\documentclass[twocolumn]{aastex62}

\usepackage{pifont}
\usepackage{float}
\usepackage{graphicx}	
\usepackage{amsmath}	
\usepackage{multirow}
\usepackage{gensymb}
\usepackage{booktabs}
\usepackage{times}
\usepackage{xcolor}
\usepackage{ulem}
\usepackage{hyperref}
\graphicspath{{./}{figures/}}
\defcitealias{perez_2020}{P20}
\defcitealias{casassus_2019}{CP19}
\accepted{August 3, 2022}

\shorttitle{Origin of the Doppler-flip in HD~100546: a large scale spiral}
\shortauthors{Norfolk et al.}

\begin{document}

\title{The Origin of the Doppler-flip in HD~100546: a large scale spiral arm generated by an inner binary companion}

\correspondingauthor{Brodie Norfolk}
\email{brodiejamesnorfolk@gmail.com}

\author[0000-0001-5898-2420]{Brodie J. Norfolk}
\affil{Centre for Astrophysics and Supercomputing (CAS), Swinburne University of Technology, Hawthorn, Victoria 3122, Australia}

\author[0000-0001-5907-5179]{Christophe Pinte}
\affil{School of Physics and Astronomy, Monash University, Vic 3800, Australia}
\affil{Univ. Grenoble Alpes, CNRS, IPAG, F-38000 Grenoble, France}

\author[0000-0001-7764-3627]{Josh Calcino}
\affil{Theoretical Division, Los Alamos National Laboratory, Los Alamos, NM 87545, USA}

\author[0000-0003-1502-4315]{Iain Hammond}
\affil{School of Physics and Astronomy, Monash University, Vic 3800, Australia}

\author{Nienke van der Marel}
\affil{Leiden Observatory, Niels Bohrweg 2, 2333 CA Leiden, The Netherlands}

\author[0000-0002-4716-4235]{Daniel J. Price}
\affil{School of Physics and Astronomy, Monash University, Vic 3800, Australia}

\author[0000-0001-5827-4088]{Sarah T. Maddison}
\affil{Centre for Astrophysics and Supercomputing (CAS), Swinburne University of Technology, Hawthorn, Victoria 3122, Australia}

\author[0000-0002-0101-8814]{Valentin Christiaens}
\affil{Space sciences, Technologies \& Astrophysics Research (STAR) Institute, Universit\'e de Li\`ege, All\'ee du Six Ao\^ut 19c, B-4000 Sart Tilman, Belgium}

\author[0000-0001-9423-6062]{Jean-Fran\c{c}ois Gonzalez}
\affil{Univ Lyon, Univ Claude Bernard Lyon 1, ENS de Lyon, CNRS, Centre de Recherche Astrophysique de Lyon UMR5574, F-69230, Saint-Genis-Laval, France}

\author{Dori Blakely}
\affil{University of Victoria, 3800 Finnerty Rd, Victoria, BC, V8P 5C2, Canada}

\author[0000-0003-4853-5736]{Giovanni Rosotti}
\affil{Leiden Observatory, Niels Bohrweg 2, 2333 CA Leiden, The Netherlands}
\affil{School of Physics and Astronomy, University of Leicester, Leicester, LE1 7RH, United Kingdom}

\author[0000-0002-4438-1971]{Christian Ginski}
\affil{Leiden Observatory, Niels Bohrweg 2, 2333 CA Leiden, The Netherlands}

\begin{abstract}
Companions at sub-arcsecond separation from young stars are difficult to image. However their presence can be inferred from the perturbations they create in the dust and gas of protoplanetary disks. Here we present a new interpretation of SPHERE polarised observations that reveal the previously detected inner spiral in the disk of HD~100546. The spiral coincides with a newly detected $^{12}$CO inner spiral and the previously reported CO emission Doppler-flip, which has been interpreted as the signature of an embedded protoplanet. Comparisons with hydrodynamical models indicate that this Doppler-flip is instead the kinematic counterpart of the spiral, which is likely generated by an inner companion inside the disk cavity.
\end{abstract}


\section{Introduction} \label{sec:intro}
High resolution observations from ALMA and VLT SPHERE have revealed that numerous transition disks contain spirals that propagate in altitude from millimeter grains in the mid-plane up to micro-meter grains in the upper layers of the disk \citep[e.g. V1247 Ori, HD135355B, MWC 758, HD100453; ][]{2017ApJ...848L..11K, 2018A&A...619A.161C, 2018ApJ...860..124D, Rosotti2020}. Observing these spirals, and spiral induced features (e.g. non-Keplerian motion) in disks \citep[for \(\rm M_{disk} \leq 10\% M_{star}\)][]{toomre_1964} presents strong evidence for the presence of a disk-perturbing companion \citep{dong_2015, Pinte2018, Calcino2021}.
\section{Methods}
\subsection{IRDIS polarimetric observations}

\begin{figure*}
	\includegraphics[width=0.999\textwidth]{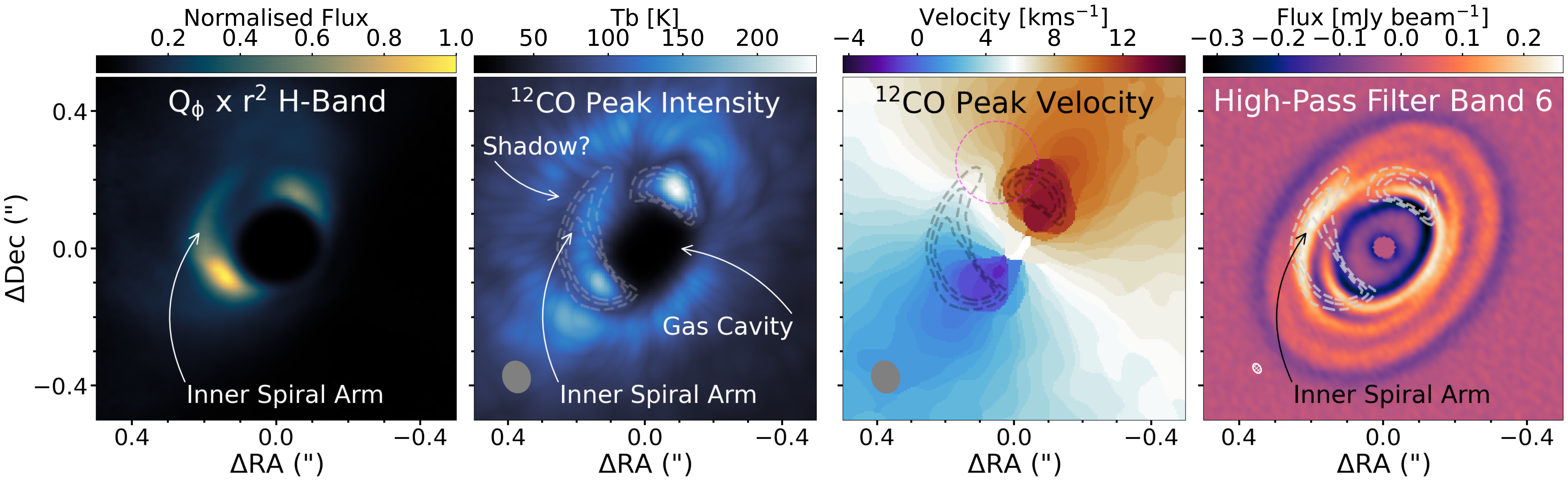}
	\caption{Panel 1: $Q_\phi$ $r^2$-scaled \textit{H}-Band observations with a sinh colorbar to emphasize bright emission. Panel 2: Inner region of the \(\rm ^{12}\)CO non-continuum subtracted peak intensity map, a sinh colorbar is used to emphasize the bright inner gas spiral. Panel 3: Peak velocity map of continuum subtracted \(\rm ^{12}\)CO emission. The magenta dashed circle represents the location of the Doppler flip reported by \citetalias{casassus_2019}. Panel 4: Our Band 6 continuum observations with a high-pass filter, we masked the bright unresolved inner disc to highlight the emission structure in the ring component, see Appendix \ref{sect:app_a} for further details. Over-plotted in each panel are the black/white dashed contours from the SPHERE observations at 0.53, 0.61, and 0.69 normalised flux units.}
	\label{fig:sphere_overlay}
\end{figure*}

The disk surrounding HD100546 \citep[distance: \(\rm 108.12\pm 0.44~pc\), age: \(\rm 5.5^{+1.4}_{-0.8}\)~Myr, mass: \(\rm 2.05^{+0.10}_{-0.12}\) \(\rm M_{\odot}\);][]{gaia_dr3, vioque_2018} contains a large spiral arm in the inner disk, and spirals at larger scales as seen in scattered light observations \citep{2016A&A...588A...8G, Folette2017, Sissa2018}. It also harbors a Doppler-flip, \emph{i.e.} a sign reversal in the $^{12}$CO rotation map at \(\rm \sim \)20~au from the star where the Keplerian rotation has been subtracted
\citep[][hereafter \citetalias{casassus_2019, perez_2020}]{casassus_2019, perez_2020}. \citetalias{casassus_2019} and \citetalias{perez_2020} suggested that these deviations from Keplerian rotation might be the imprint of a $>5$\,M$_\mathrm{Jup}$ embedded protoplanet, where the Doppler-flip arises from the opposite signs of the velocity field, relative to the planet, in the inner and outer Lindblad resonance spirals. 

However, as discussed by the authors, the planet origin for the non-Keplerian motions is difficult to reconcile with the sub-millimeter continuum observations. The Doppler-flip, and hence the tentative planet, are co-located with the azimuthal asymmetry seen in the dust continuum ring, while massive embedded planets are expected to carve gaps in the dust (and gas) distribution \citep[e.g.][]{paardekooper_2004}.

In this letter, we compare SPHERE $H$-band scattered light images with ALMA $^{12}$CO J=2-1 observations, and present hydrodynamical and radiative transfer models to explore whether the disk central cavity (\(\rm \sim 16.5\)~au) and CO kinematic structure share a common origin, namely a close companion carving the cavity and generating spiral arms.

HD100546 was observed with the Infrared Differential Imaging Spectrometer \citep[IRDIS;][]{Dohlen:2008uh} instrument on the  Spectro-Polarimetic High contrast imager for Exoplanets REsearch \citep[SPHERE;][]{Beuzit2008}, on
18 February 2019 as part of ESO programme 0102.C-0162(A) (PI: C. Ginski). We used the dual-polarisation imaging (PDI) mode \citep{de-Boer:2020vx}, with an integration time of 16 s, a \textit{H}-band filter (1.625~µm) and an apodised Lyot coronagraph ALC2 (\texttt{N\_ALC\_YJH\_S}, diameter 185 mas).
We used eight polarimetric cycles with two exposures at each half wave plate position (total integration time 17 minutes) with an average seeing of 0\rlap{.}\arcsec82 (after excluding the first polarimetric cycle and several cycles that showed inconsistent coronographic status).

We reduced the archival observations (unpublished) using the IRDIS Data reduction for Accurate Polarimetry pipeline \citep[\texttt{IRDAP};][]{van-Holstein:2020tm}.
We computed the clean \textit{Q} and \textit{U} Stokes parameters using the normalised double-difference method, and corrected for instrument polarisation by applying the Mueller matrix model. A \textit{Q\textsubscript{$\phi$}} $r^2$-scaled image was produced using \texttt{diskmap} \citep{Stolker:2016ub} to account for stellocentric flux loss to 150~au, using the revised \textit{Gaia} distance of  $\rm 110\pm0.6$~pc, and disk inclination 43.2\degr~and PA 144\degr~retrieved from Markov chain Monte Carlo fits of the sub-millimeter continuum emission.


\subsection{ALMA Observations and Data Reduction}
\begin{figure*}
	\includegraphics[width=0.99\textwidth]{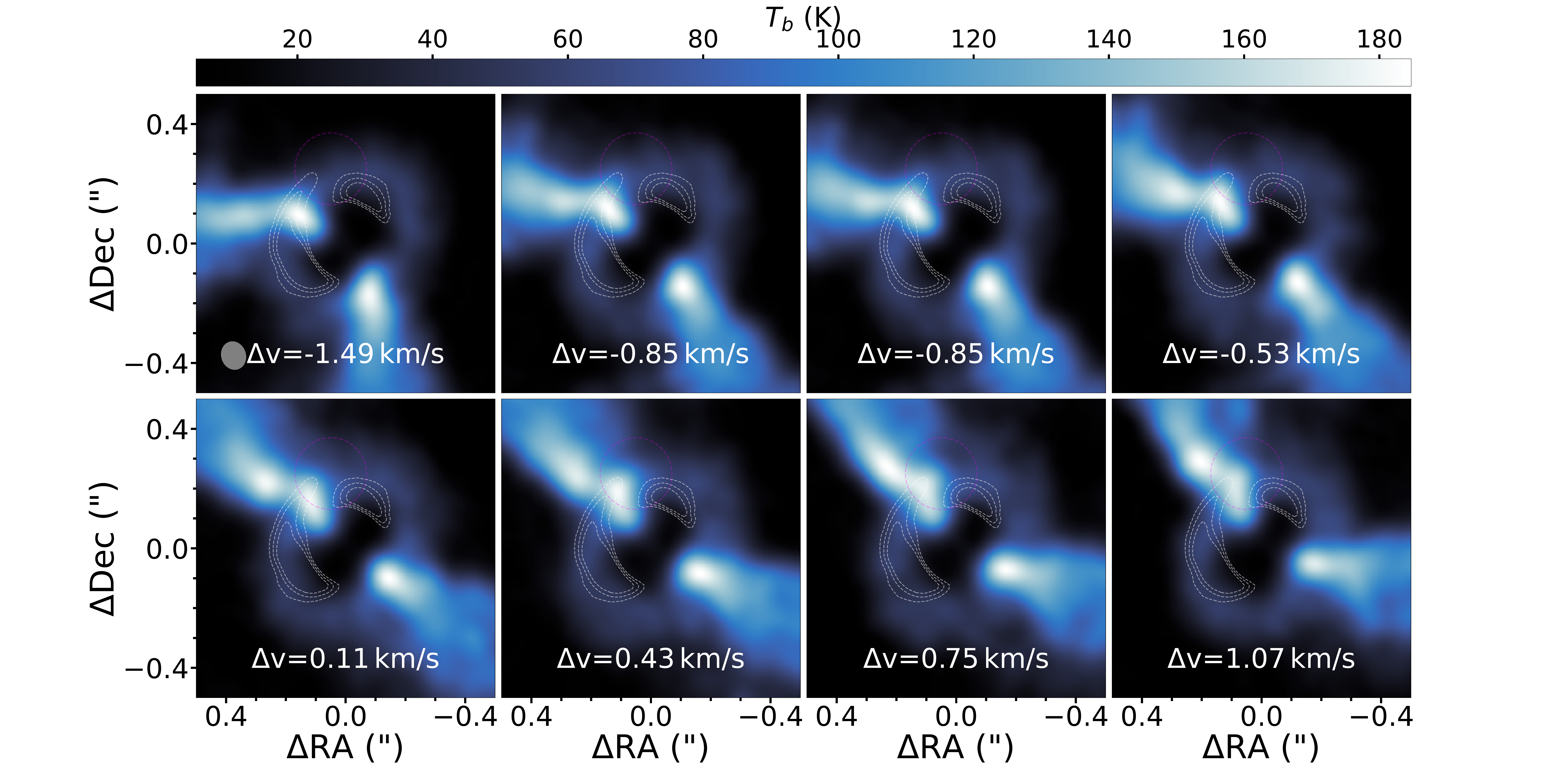}
	\centering
	\caption{Selected non-continuum subtracted $^{12}$CO J=2-1 channel maps highlighting the deviation from Keplerian velocity at the edge of the disk cavity. The lower limit on the colorbar is set to 15~K. The SPHERE $Q_{\phi}$ image is shown with white dashed contours at 1.3e6, 1.5e6, and 1.7e6 adu. White arrows point to the "kink" as it shifts across each channel map. The magenta dashed circle represents the location of the Doppler flip reported by \citetalias{casassus_2019}.}
	\label{fig:channel_maps}
\end{figure*}

We calibrated archival ALMA Band 6 observations (Projects: 2016.1.00344.S, published in \citetalias{casassus_2019} and \citetalias{perez_2020}; and 2018.1.01309.S, published in \citealp{casassus_2022}) using the {\sc CASA} pipeline for the appropriate ALMA cycle.
Continuum observations have a total bandwidth of 2~GHz for 2016.1.00344.S and 6~GHz for 2018.1.01309.S. The $^{13}$CO and C$^{18}$O J=2-1 lines were observed with a resolution of 122~kHz in the 2016.1.00344.S program, and the $^{12}$CO J=2-1 line was observed at 122~kHz resolution in both programs. We performed four rounds of phase-only self-calibration on the continuum short baseline data of 2016.1.00344.S with successive integration times of  \textit{`inf'}, 120s, 60s, and  \textit{`int'}. For the long baseline continuum data of both programs, we performed a single phase-only self-calibration with an infinite integration time.
We shifted each execution to a common position 
with the \textit{`fixvis'} task, and re-scaled the flux of the long baseline visibilities to match the short baseline data using the DSHARP reduction utilities, prior to concatenating all the executions. We applied the same self-calibration solutions, shift and re-scaling to the CO lines.


The continuum and CO emission
were imaged
using the \textit{tclean} task in CASA, with the Multiscale Clean deconvolver \citep{2008ISTSP...2..793C} with Briggs weighting. We used a  robust parameter of 0.5 for the continuum, leading to a beam size of 0.03\arcsec $\times$ 0.02\arcsec, with a PA of 31$^\circ$, a SNR of 320 and RMS of 9.8 \(\rm \mu\)Jy beam$^{-1}$.
For the $^{12}$CO emission (with and without continuum subtraction) we used a channel width of 0.32~\(\rm kms^{-1}\), a robust parameter of 1.0, and a $uv$ taper of 0.06\arcsec, resulting in a beam size of 0.95\arcsec $\times$ 0.081\arcsec, with a PA of 16$^\circ$ and a RMS noise of 0.8 mJy \(\rm beam^{-1}\) for our continuum subtracted cube, and beam size 0.094\arcsec $\times$ 0.081\arcsec, PA of 18$^\circ$ and RMS noise of 1.0 mJy \(\rm beam^{-1}\) for our non-subtracted cube. We then applied JvM \citep{jvm1995} and primary beam corrections to all images.
For $^{13}$CO and C$^{18}$O emission (continuum subtracted) we used a channel width of 0.32~\(\rm kms^{-1}\), a robust parameter of 1.0, and a $uv$ taper of 0.06\arcsec. This resulted in a beam size of 0.12\arcsec $\times$ 0.10\arcsec, with a PA of -20$^\circ$ and a RMS noise of 1.7 mJy \(\rm beam^{-1}\) for $^{13}$CO, and a beam size of 0.12\arcsec $\times$ 0.10\arcsec, with a PA of -19$^\circ$ and a RMS noise of 1.3 mJy \(\rm beam^{-1}\) for C$^{18}$O.

\subsection{Hydrodynamics and radiative transfer modeling}

We performed 3D smoothed particle hydrodynamics (SPH) simulations using \textsc{phantom} \citep{Price2018b}. Our goal was not to perform a detailed fitting of all the available observations, but to explore if a binary can explain both the central cavity and disrupted kinematics. We assumed a central star with mass 2 M$_\odot$ \citep{vioque_2018}. Based on previous work on synthetic observations of circumbinary disks \citep{Ragusa2017,Price2018b}, we only explored a restricted region of the parameter space with companion masses between 0.2 and 0.6 M$_\odot$ (q=0.1 and 0.3) at a semi-major axis of 8.4~au \citep[e.g. \(\rm \sim 1/2 \times\) the cavity radius of \(\rm \sim 16.2\)~au,][]{artymowicz_1994} and on orbits with eccentricities between 0.2 and 0.6. We treated both stars as sink particles \citep{bate1995} with accretion radii of 0.7~au. We initialized a gas-only circumbinary disk with \(\rm 2\times\)1\(\rm 0^6\) SPH particles between $R_\textrm{in} = 16.8$~au and $R_\textrm{out} = 126$~au, with a surface density profile $\Sigma (r) \propto r^{-1}$, a temperature profile $T \propto r^{-0.5}$, and a scale height $H/R_\textrm{ref} = 0.05$ at $R_\textrm{ref} = 25.2$~au (chosen to closely match the scale height of the disc, see Appendix \ref{sect:app_c}, where \(\rm (H/R)_{^{12}CO} \approx 3-4 \times H/R_{ref}\)). We used a total gas disk mass of 5$\times$10$^{-3}$\,M$_\odot$. Given this low mass we neglect the disc self-gravity. We modeled the disc viscosity using an average \cite{shakura1973} viscosity of $\alpha_{SS} = 5\times 10^{-3}$ by setting the SPH artificial viscosity $\alpha_{AV} = 0.24$.

Length values are given after a re-scaling of all length (and consequentially velocity and time) code units in our \textsc{phantom} models to match the models cavity radius with the observed cavity radius \citep[e.g. \(\rm \sim\) 16~au,][]{perez_2020}.

We ran our models for 100 orbits of the central binary, and then re-ran them for one additional orbit with outputs every 1/10 of an orbit to study how the disk structure varies as a function of the binary phase. Although the duration of our simulation is lower than the viscous timescale, circumbinary disc simulations by \cite{hirsh2020} show that the cavity size is largely set on timescales of a few hundred binary orbits.

For each of the 10 output samples along the final orbit of our SPH simulations, we generated a Voronoi mesh for input into the Monte Carlo radiative transfer code \textsc{mcfost} \citep{pinte2006, pinte2009} which is used to produce synthetic CO emission observations. We assumed a power-law dust grain size distribution $\rm dn/ds \propto s^{-3.5}$ for $0.05\ \mu\textrm{m} \leq s \leq 3000 \ \mu\textrm{m} $ with a gas-to-dust ratio of 20 (\citet{Bruderer2012} suggests a ratio of 100 is a better fit to their \textit{Herschel} data however, their alternative solution of $\rm Gas/Dust = 20$ is in closer agreement with more recent derivations using ALMA data \citep{miley_2019}), for a population of spherical and homogeneous grains composed of astronomical silicate \citep{weingartner2001}. We set $\rm T_eff=10250$ and $R_{\odot}=1.91$ for HD~100546 as derived by \citet{vioque_2018}. We assumed that our companion is approximately the same age as its host \citep[e.g. 5~Myr,][]{vioque_2018} and used the isochrones from \cite{Siess2000}, which led to and effective temperatures of 3870\,K and a radii of 1.16 R$_\odot$.
We used an initial CO abundance of CO/H$_2 = 1\times 10^{-4}$. This ratio is affected by freeze out when T $<$ 20~K (depletion factor of $1\times 10^{-4}$), full photo-dissociation by UV radiation, as well as photo-desorption of the CO ice (following appendix B of \citealp{Pinte2018}). We generated channel maps with a spectral resolution of 0.32 km/s and convolved them by the observed beam. The resulting synthetic cubes do not reproduce the observed peak temperature ($\sim$130~K vs. $\sim$180~K) however, this likely indicates that $\rm T_{gas} \geq T_{dust}$ in the CO emitting layer, due to for instance to PAH photoelectric heating \citep{woitke_2009}, which is not included here. A similar difference was observed by \cite{de_gregorio_monsalvo_2013} for HD~163296.

\begin{figure*}
	\includegraphics[width=\textwidth]{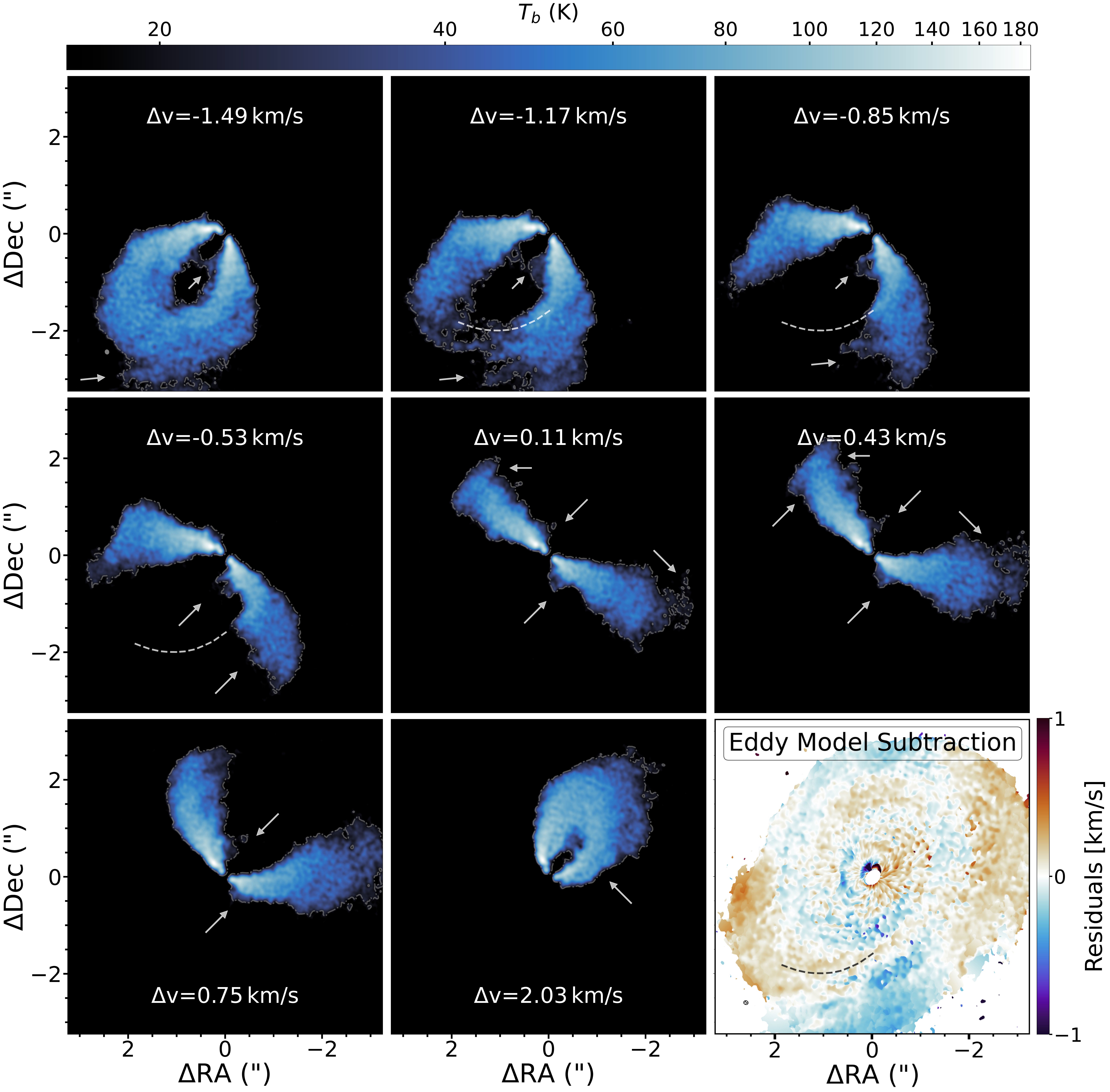}
	\centering
	\caption{Large scale continuum subtracted CO(2-1) channel maps. We highlight several non-Keplerian features with white arrows and overlay a trace of the southern scattered light spiral (see Fig.~\ref{fig:qphi_m8}) with a dashed line. We include a large scale view of our \textsc{eddy} model subtraction residuals, and highlight with a dashed black contour that a prominent velocity spiral arm in the residual map also coincides with the scattered light spiral. To ensure these deviations are not background noise, we plot the 5$\rm \sigma$ value ($\rm 5 \times $1.2 mJy \(\rm beam^{-1}\)) with a dash-dot gray contour in each channel.}
	\label{fig:channel_maps_large_scale}
\end{figure*}

\section{Results and Discussion}
\label{sect:results}

\subsection{The Doppler-flip is the kinematic counterpart of a spiral arm}

We re-detect a large scale spiral in scattered light (Figure \ref{fig:sphere_overlay}, left panel) originating from the inner edge of the disk ($\approx$ 0.12\arcsec) in the South-East direction and opening to the North up to $\approx$ 0.25\arcsec separation \citep[previously imaged by][]{2016A&A...588A...8G, Folette2017, Sissa2018}. We also detect this inner spiral arm in non-continuum subtracted $^{12}$CO peak brightness temperature (second panel), and a tentative shadow behind the spiral (both the spiral and shadow are detected in the continuum subtracted cube albeit at a lesser extent). The scattered light spiral aligns with this newly detected gas spiral as well as the bending of the isovelocity curves in the $^{12}$CO peak velocity map (third panel), \emph{i.e.} with what appears as the Doppler-flip in the $\rm v_{Keplerian}$ subtracted rotation map in \citetalias{casassus_2019} (Fig.~1). Equivalently, the scattered light spiral is co-located with the velocity kinks observed in the channel maps (Fig.~\ref{fig:channel_maps}), which is also tentatively seen in $^{13}$CO and C$^{18}$O emission (e.g. see Appendix \ref{sect:app_e}).

Since scattered light and $^{12}$CO emission arise from similar altitudes in the disk (e.g. see Appendix \ref{sect:app_c}), projection effects should be minimal, which suggests that the scattered light spiral is likely to be physically connected to the non-Keplerian motions observed in $^{12}$CO.

We observe evidence of additional co-spatial features in the disk. Figure~\ref{fig:channel_maps_large_scale} highlights several deviations from Keplerian motion at large scales in the $^{12}$CO channel maps, which includes a large kink in the \(\rm \Delta v = -0.85ms^{-1}\) and \(\rm 0.53kms^{-1}\) channels, also visible in our \textsc{eddy} model subtraction residuals, that coincide with the outer scattered light spiral arm (this connection is also detected in our $^{12}$CO peak intensity map, see Appendix~\ref{sect:app_b} for further details). Additionally, using a high band-pass filter on the 1.3~mm dust continuum observations, we trace a tentative spiral structure (Fig.~\ref{fig:sphere_overlay}, right panel) crossing the dust crescent at the inner edge of the cavity. Assuming sufficient dust settling and a relatively lower optical depth at 1.3~mm, this spiral structure is likely to be the mid-plane counterpart of the scattered light spiral \citep[as similarly found for HD100453,][]{Rosotti2020}.

\subsection{An inner binary generates an apparent Doppler-flip}

\begin{figure*}
  \includegraphics[width=\textwidth]{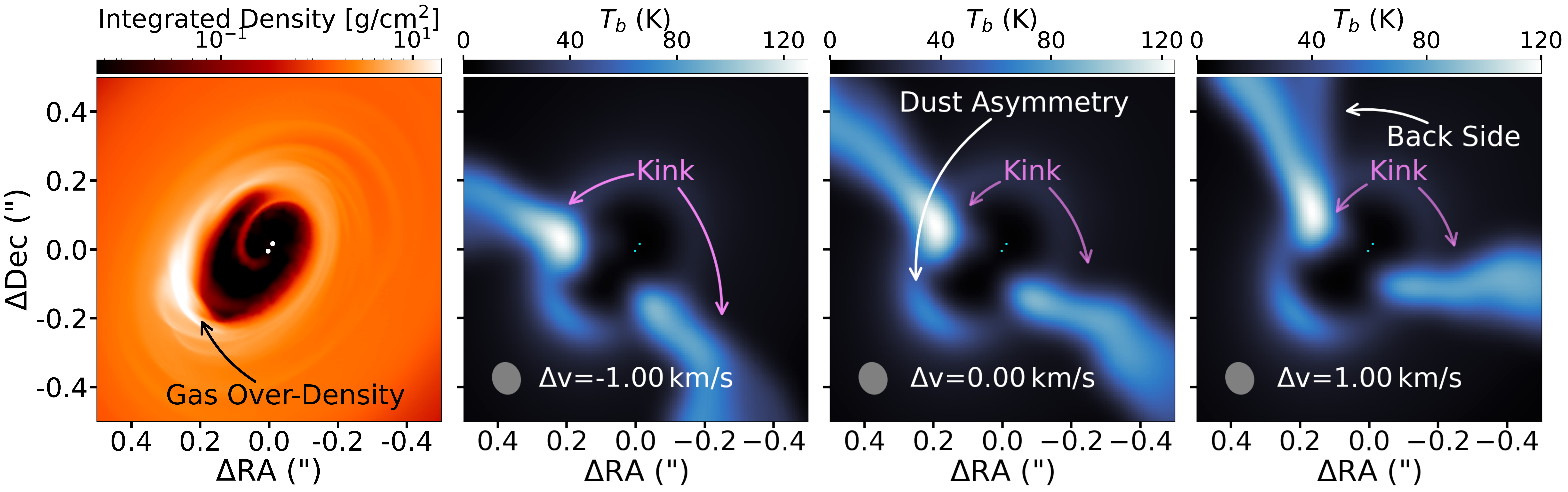}
  \caption{Panel 1: Integrated density from our SPH simulation. Panels 2, 3 \& 4: Synthetic channel maps from our circumbinary model with a $\sim 0.6$ M$_\odot$ ($q=0.3$) companion at 8.4~au convolved with the same beam dimensions as the observations.
  \label{fig:model_channels}}
\end{figure*}

Our modeling shows that a planet at \(\rm \sim \)20~au is not required to produce a kinematic structure that resembles the Doppler-flip detected in HD~100546. 
The surface density of the SPH simulation with $q=0.3$ ($\rm M_{binary}/M_{star}$) and $e=0.6$ (eccentricity) (our fiducial model) along with the corresponding $^{12}$CO channel maps are shown in Figure \ref{fig:model_channels}. The central binary is able to produce the prominent spiral arms, with the associated non-Keplerian motions. Our model displays several velocity kinks at the edge of the cavity, with amplitude and shape similar to those observed in HD~100546. As out goal is only to qualitatively explain the observations, we explored a limited region of the parameter space, and the selected model is not a unique solution. The amplitude of the kinematic deviation increases with mass ratio and eccentricity. Models with q$\leq$0.2 did not perturb the disc sufficiently to produce the observed non-Keplerian motion, suggesting a stellar mass companion was required.We note that a more massive companion might have led to an even better agreement with the observations, but a mass ratio of 0.3 is already in tension with the constraints of sparse aperture masking (see discussion in section \ref{ref:companion_mass}), so we did not explore higher masses. 

\begin{figure*}
	\includegraphics[width=\textwidth]{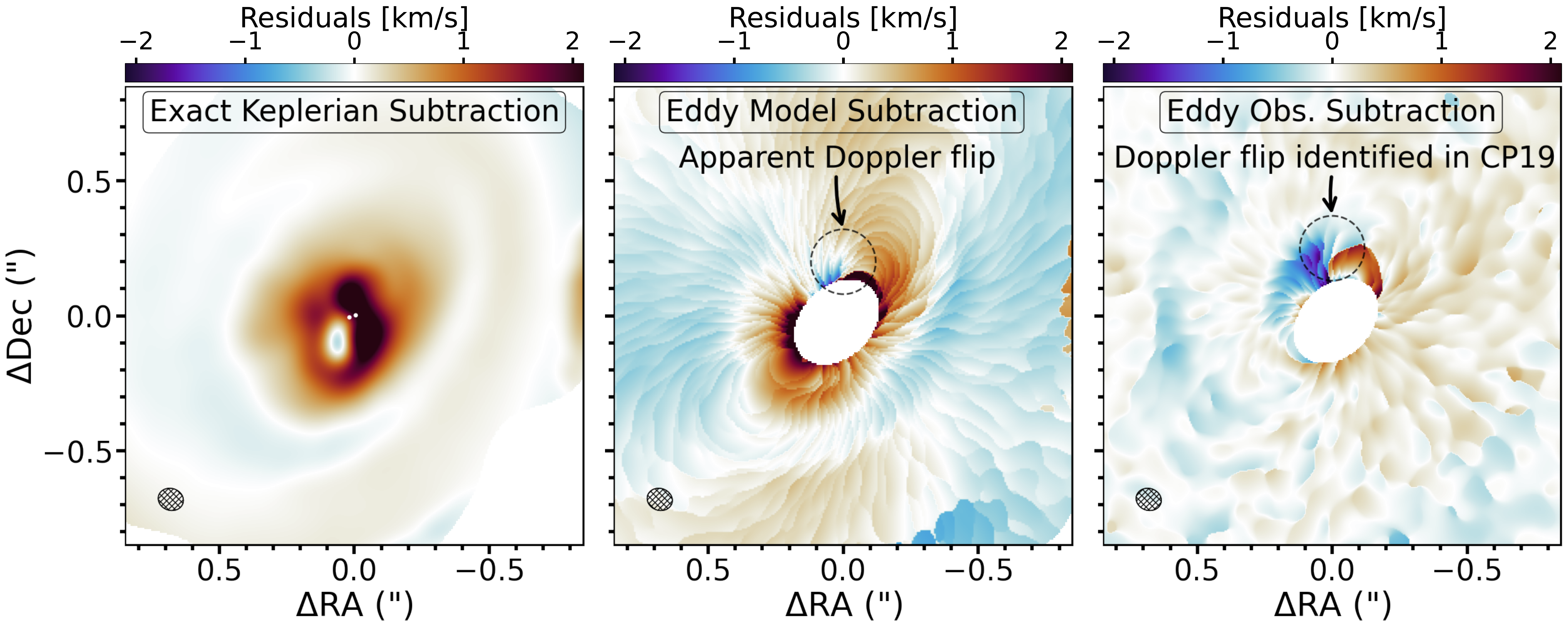}
	\centering
	\caption{ Left: Residuals from the known Keplerian velocities, \emph{i.e.} by subtracting a model where we force the gas velocities to be circular and Keplerian around the center of gravity of the model, assuming a central mass of 2.6\,M$_\odot$ (i.e. the total mass of the binary.) Center: Best-fit Keplerian rotation subtracted with {\sf eddy} from the moment 1 of our SPH simulation. Right: As in the center panel, but for the continuum subtracted $^{12}$CO observations of HD~100546 presented in this work. The binary induced spiral arms in our SPH simulation produce substantial deviations ($\sim 1$ km/s) from Keplerian rotation over short distances, which appears like a Doppler-flip when trying to subtract a estimated velocity field. This is similar to the deviations seen in HD~100546 by \citetalias{casassus_2019}.}
	\label{fig:model_flip}
\end{figure*}

To assess if the fiducial model's kinematic structure can appear as a Doppler-flip, we subtracted an ``exact" Keplerian velocity field from the model peak velocity map (by forcing the SPH particles to be on a circular Keplerian orbit around a central object with a mass equal to the sum of the 2 binary objects), and present the residuals in Fig.~\ref{fig:model_flip} panel 1 (adopting a procedure similar to that used in \citealp{Pinte2020}). The resulting velocity residuals reveal that the gas flow is strongly non-Keplerian, with velocity spirals at multiple scales, but do not show a Doppler-flip at the edge of the cavity.

However, for the observed disk, the unperturbed velocity field  is unknown, and is instead estimated either by fitting a model \citep[e.g.][]{Teague2018} or by constructing an azimuthal average \citepalias{casassus_2019}. We then perform a similar procedure on our synthetic observations, as well as on the observations (Fig.~\ref{fig:model_flip}, central and right panels). We generate a peak velocity map with \textsc{bettermoments} \citep{bettermoments2018}, before subtracting a Keplerian model using \textsc{eddy}  \citep{eddy}. Both the synthetic and observed cubes are processed in the same way. 

The right panel of Fig.~\ref{fig:model_flip} shows a sign reversal in the rotation map that is consistent with the Doppler-flip detected by \citetalias{casassus_2019}. Interestingly, we detect a similar flip in the synthetic observations (central panel) and spiral-like residuals. A comparison with the left panel shows that this apparent Doppler-flip is an artifact of the subtraction procedure. \textsc{eddy} tries to subtract a Keplerian rotation field that best matches the observations. This relies on the assumption that a smooth Keplerian rotation background dominates the velocity field. Consequentially for our synthetic residual map, \textsc{eddy} slightly underestimates the central mass (2.4 instead of 2.6\,M$_\odot$). The subtraction of this estimated velocity average from the strongly modulated rotation map then results in an apparent sign reversal in the residuals. 
\subsection{Where is the companion ?}
\label{ref:companion_mass}

The similarity between the \textsc{eddy} residuals for the $^{12}$CO data and synthetic observations from the SPH simulation suggests that the Doppler-flip detected by \citetalias{casassus_2019} may simply be the summation of 'kinks' across the systematic velocity that remain after subtracting an average velocity background. We showed that the observed kinematic structures in HD~100546 \citep[apparent Doppler-flip, non-Keplerian motions in the cavity, \citetalias{perez_2020}, initially reported as a misaligned inner gas disk;][]{2017A&A...607A.114W} can be explained by a central binary, rather than a planet located inside the dust ring. A central binary can also explain the presence of many of the structures observed by other disc tracers such as: the large cavity and asymmetry observed in continuum wavelengths \citep{perez_2020, 2021MNRAS.502.5779N}, the spiral arm seen in scattered light \citep[][and see our model comparison in Appendix \ref{sect:app_d}]{2016A&A...588A...8G, Folette2017, Sissa2018}, as well as the non-Keplarian motions observed in the inner regions of the disc via SO emission \citep{booth_2018}. A single binary companion has been shown to drive similar disk features in both HD142527 \citep{Price2018} and potentially IRS~48 \citep{2019MNRAS.490.2579C}, although a companion has not been detected directly in the latter.

A series of observational studies have suggested the presence of a companion inside the cavity of HD~100546 \citep{Brittain2014, Brittain2019}, although most have discussed a several-Jupiter mass planet rather than a stellar mass companion. \citetalias{perez_2020} conducted sparse aperture masking observations (SAM) with SPHERE and concluded that no stellar companion exists in the cavity. 
Our model predicts a apparent separation of 25mas, and a flux ratio of 8.4 ($\Delta M$ = 2.3) in K band between the two stars. According to the detection limits of the SPHERE SAM observations (Fig.~6 in \citetalias{perez_2020}), such a companion should have been detected. This discrepancy might be resolved with a lower mass companion that is inclined relative to the outer disk. Inclined lower mass companions will generate non-Keplerian motions at similar magnitudes \citep[e.g. as was seen for the variety of inclinations used for HD~142527b's orbit,][]{Price2018} while remaining undetected in the SAM observations. However, we did not consider inclined configurations in this study for simplicity. 
We also note that while the Gravity data reveals a large fractional contribution of the circumstellar disk in the VLTI field of view: 0.63 in $K$ band, the visibilities and phase closures display high-frequency modulations \citep[Fig. B.1 in][]{Gravity2019} that may indicate the presence of a companion.

\cite{gaia_dr3} indicates a preliminary astrometric excess noise of 0.38~mas at the 155\,$\sigma$ level, which is significantly larger that the 0.16~mas of HD~142527, where there is a known $\approx$ 0.2~M$_\odot$ binary companion \citep{Price2018b}. While the astrometric signal in a young star can be perturbed by variability and illumination effects, making it difficult to definitely attribute any excess noise to a physical astrometric wobble caused by a companion, the large excess in HD~100546 does suggest the presence of a binary companion.

The short orbital period of a potential binary (about 15 years in our model) also offers some exciting opportunities for monitoring to test our interpretation. For instance, our model predicts that for most of the binary orbit (7/10 dumps across 1 orbit), the velocity kink is flipped (at the same azimuthal position) compared to the current configuration. If our prediction is correct, kinematic signatures should change significantly over the next 3 to 5 years, and regular ALMA follow-up observations could test the validity of our proposed binary-disk configuration. If the orbit is indeed eccentric, additional observations with SAM might also be able to reveal the companion within a few years.

Although a companion inside the cavity can explain the morphological features around the mm dust cavity, an additional mechanism might be required to explain features at large scales. A dust ring with radius $\sim 180$ au was reported by \cite{Fedele2021}, and is potentially created by a massive planet at $\sim 80 - 120 $ au. Such a planet could produce morphological and kinematic signatures such as localised kinks \citep[e.g.][]{Pinte2018} or more extended velocity perturbations from spiral arms \citep{Calcino2021}. Another possibility is that the disk around HD100546 is being perturbed by late stage inflows \citep{dullemond2019}. Either or both of these scenarios could explain the spiral arms and kinematic perturbations seen at distances larger than 100\,au ($\gtrsim 1$\arcsec) from the central cavity in Figures \ref{fig:channel_maps_large_scale} and \ref{fig:qphi_m8}.  Interestingly, this is similar to HD 142527, where \cite{Garg2021} detected large scale spirals up to 300\,au while the central binary semi-major axis ranges from 12 to 31\,au \citep{Claudi2019}.

\section{Summary}
\begin{enumerate}
\item Using archival SPHERE observations, we confirmed the prominent spiral arm stemming from the edge of the cavity, previously detected \citep{2016A&A...588A...8G, Folette2017, Sissa2018}, and note that it spatially aligns with the `kink' observed in the \(\rm ^{12}\)CO channel maps, or equivalently with the Doppler-flip seen in the rotation map residuals.
\item Our hydrodynamical modeling shows that  a $\sim 0.6$~M$_\odot$ ($q=0.3$) companion at $\approx$8 au can reproduce many of the observed disc features including: an inner $\rm ^{12}$CO cavity, a velocity kink and corresponding residual Doppler flip (albeit at a weaker magnitudes) at the edge of the cavity, a scattered light spiral, and a cavity and asymmetry in the millimeter dust.
\item We detect several new non-Keplerian features at large scales across the whole disk and at all velocities, some of which also appear to spatially coincide with spirals seen in scattered light, the origin of which remains unclear.
\end{enumerate}

\acknowledgments

B.J.N. is supported by an Australian Government Research Training Program Scholarship. C.P., D.P. acknowledges funding from the Australian Research Council via FT170100040 and DP180104235. J.-F.G. acknowledges funding from ANR (Agence Nationale de la Recherche) of France under contract number ANR-16-CE31-0013 (Planet-Forming-Disks). V.C. acknowledges funding from the Belgian F.R.S.-FNRS. This project has received funding from the European Union's Horizon 2020 research and innovation programme under the Marie Sk\l{}odowska-Curie grant agreements No 210021 and No 823823 (DUSTBUSTERS). GR acknowledges support from the Netherlands Organisation for Scientific Research (NWO, program number 016.Veni.192.233) and from an STFC Ernest Rutherford Fellowship (grant number ST/T003855/1). SPH simulations were performed on OzStar, funded by Swinburne University of Technology and the Australian government. The National Radio Astronomy Observatory is a facility of the National Science Foundation operated under agreement by the Associated Universities, Inc. ALMA is a partnership of ESO (representing its member states), NSF (USA) and NINS (Japan), together with NRC (Canada) and NSC and ASIAA (Taiwan) and KASI (Republic of Korea), in cooperation with the Republic of Chile. The Joint ALMA Observatory is operated by ESO, AUI/ NRAO and NAOJ. This work has made use of data from the European Space Agency (ESA) mission \textit{Gaia} (\url{https://www.cosmos.esa.int/gaia}), processed by the \textit{Gaia} Data Processing and Analysis Consortium (DPAC; \url{https://www.cosmos.esa.int/web/gaia/dpac/consortium}). Funding for the DPAC has been provided by national institutions, in particular the institutions participating in the \textit{Gaia} Multilateral Agreement.

%

\vspace{5mm}
\facilities{Atacama Large millimeter/sub-millimeter Array (ALMA), Spectro-Polarimetic High contrast imager for Exoplanets REsearch (SPHERE).}


\software{CASA \citep{2007ASPC..376..127M},
IRDAP \citep{van-Holstein:2020tm},
PHANTOM \citep{Price2018},
MCFOST \citep{pinte2006, pinte2009}
}

\bibliography{paper}{}

\begin{thebibliography}{}
\expandafter\ifx\csname natexlab\endcsname\relax\def\natexlab#1{#1}\fi
\providecommand{\url}[1]{\href{#1}{#1}}

\bibitem[{{Artymowicz} \& {Lubow}(1994)}]{artymowicz_1994}
{Artymowicz}, P., \& {Lubow}, S.~H. 1994, \apj, 421, 651

\bibitem[{{Bate} {et~al.}(1995){Bate}, {Bonnell}, \& {Price}}]{bate1995}
{Bate}, M.~R., {Bonnell}, I.~A., \& {Price}, N.~M. 1995, \mnras, 277, 362

\bibitem[{{Beuzit} {et~al.}(2008){Beuzit}, {Feldt}, {Dohlen}, {Mouillet},
  {Puget}, {Wildi}, {Abe}, {Antichi}, {Baruffolo}, {Baudoz}, {Boccaletti},
  {Carbillet}, {Charton}, {Claudi}, {Downing}, {Fabron}, {Feautrier},
  {Fedrigo}, {Fusco}, {Gach}, {Gratton}, {Henning}, {Hubin}, {Joos}, {Kasper},
  {Langlois}, {Lenzen}, {Moutou}, {Pavlov}, {Petit}, {Pragt}, {Rabou}, {Rigal},
  {Roelfsema}, {Rousset}, {Saisse}, {Schmid}, {Stadler}, {Thalmann}, {Turatto},
  {Udry}, {Vakili}, \& {Waters}}]{Beuzit2008}
{Beuzit}, J.-L., {Feldt}, M., {Dohlen}, K., {et~al.} 2008, in Society of
  Photo-Optical Instrumentation Engineers (SPIE) Conference Series, Vol. 7014,
  Society of Photo-Optical Instrumentation Engineers (SPIE) Conference Series

\bibitem[{{Booth} {et~al.}(2018){Booth}, {Walsh}, {Kama}, {Loomis}, {Maud}, \&
  {Juh{\'a}sz}}]{booth_2018}
{Booth}, A.~S., {Walsh}, C., {Kama}, M., {et~al.} 2018, \aap, 611, A16

\bibitem[{{Brittain} {et~al.}(2014){Brittain}, {Carr}, {Najita}, {Quanz}, \&
  {Meyer}}]{Brittain2014}
{Brittain}, S.~D., {Carr}, J.~S., {Najita}, J.~R., {Quanz}, S.~P., \& {Meyer},
  M.~R. 2014, \apj, 791, 136

\bibitem[{{Brittain} {et~al.}(2019){Brittain}, {Najita}, \&
  {Carr}}]{Brittain2019}
{Brittain}, S.~D., {Najita}, J.~R., \& {Carr}, J.~S. 2019, \apj, 883, 37

\bibitem[{{Bruderer} {et~al.}(2012){Bruderer}, {van Dishoeck}, {Doty}, \&
  {Herczeg}}]{Bruderer2012}
{Bruderer}, S., {van Dishoeck}, E.~F., {Doty}, S.~D., \& {Herczeg}, G.~J. 2012,
  \aap, 541, A91

\bibitem[{{Calcino} {et~al.}(2021){Calcino}, {Hilder}, {Price}, {Pinte},
  {Bollati}, {Lodato}, \& {Norfolk}}]{Calcino2021}
{Calcino}, J., {Hilder}, T., {Price}, D.~J., {et~al.} 2021, arXiv e-prints,
  arXiv:2111.07416

\bibitem[{{Calcino} {et~al.}(2019){Calcino}, {Price}, {Pinte}, {van der Marel},
  {Ragusa}, {Dipierro}, {Cuello}, \& {Christiaens}}]{2019MNRAS.490.2579C}
{Calcino}, J., {Price}, D.~J., {Pinte}, C., {et~al.} 2019, \mnras, 490, 2579

\bibitem[{{Casassus} {et~al.}(2022){Casassus}, {C{\'a}rcamo}, {Hales}, {Weber},
  \& {Dent}}]{casassus_2022}
{Casassus}, S., {C{\'a}rcamo}, M., {Hales}, A., {Weber}, P., \& {Dent}, B.
  2022, \apjl, 933, L4

\bibitem[{{Casassus} \& {P{\'e}rez}(2019)}]{casassus_2019}
{Casassus}, S., \& {P{\'e}rez}, S. 2019, \apjl, 883, L41

\bibitem[{{Cazzoletti} {et~al.}(2018){Cazzoletti}, {van Dishoeck}, {Pinilla},
  {Tazzari}, {Facchini}, {van der Marel}, {Benisty}, {Garufi}, \&
  {P{\'e}rez}}]{2018A&A...619A.161C}
{Cazzoletti}, P., {van Dishoeck}, E.~F., {Pinilla}, P., {et~al.} 2018, \aap,
  619, A161

\bibitem[{{Claudi} {et~al.}(2019){Claudi}, {Maire}, {Mesa}, {Cheetham},
  {Fontanive}, {Gratton}, {Zurlo}, {Avenhaus}, {Bhowmik}, {Biller},
  {Boccaletti}, {Bonavita}, {Bonnefoy}, {Cascone}, {Chauvin}, {Delboulb{\'e}},
  {Desidera}, {D'Orazi}, {Feautrier}, {Feldt}, {Flammini Dotti}, {Girard},
  {Giro}, {Janson}, {Hagelberg}, {Keppler}, {Kopytova}, {Lacour}, {Lagrange},
  {Langlois}, {Lannier}, {Le Coroller}, {Menard}, {Messina}, {Meyer},
  {Millward}, {Olofsson}, {Pavlov}, {Peretti}, {Perrot}, {Pinte}, {Pragt},
  {Ramos}, {Rochat}, {Rodet}, {Roelfsema}, {Rouan}, {Salter}, {Schmidt},
  {Sissa}, {Thebault}, {Udry}, \& {Vigan}}]{Claudi2019}
{Claudi}, R., {Maire}, A.~L., {Mesa}, D., {et~al.} 2019, \aap, 622, A96

\bibitem[{{Cornwell}(2008)}]{2008ISTSP...2..793C}
{Cornwell}, T.~J. 2008, IEEE Journal of Selected Topics in Signal Processing,
  2, 793

\bibitem[{{de Boer} {et~al.}(2020){de Boer}, {Langlois}, {van Holstein},
  {Girard}, {Mouillet}, {Vigan}, {Dohlen}, {Snik}, {Keller}, {Ginski}, {Stam},
  {Milli}, {Wahhaj}, {Kasper}, {Schmid}, {Rabou}, {Gluck}, {Hugot}, {Perret},
  {Martinez}, {Weber}, {Pragt}, {Sauvage}, {Boccaletti}, {Le Coroller},
  {Dominik}, {Henning}, {Lagadec}, {M{\'e}nard}, {Turatto}, {Udry}, {Chauvin},
  {Feldt}, \& {Beuzit}}]{de-Boer:2020vx}
{de Boer}, J., {Langlois}, M., {van Holstein}, R.~G., {et~al.} 2020, \aap, 633,
  A63

\bibitem[{{de Gregorio-Monsalvo} {et~al.}(2013){de Gregorio-Monsalvo},
  {M{\'e}nard}, {Dent}, {Pinte}, {L{\'o}pez}, {Klaassen}, {Hales},
  {Cort{\'e}s}, {Rawlings}, {Tachihara}, {Testi}, {Takahashi}, {Chapillon},
  {Mathews}, {Juhasz}, {Akiyama}, {Higuchi}, {Saito}, {Nyman}, {Phillips},
  {Rod{\'o}n}, {Corder}, \& {Van Kempen}}]{de_gregorio_monsalvo_2013}
{de Gregorio-Monsalvo}, I., {M{\'e}nard}, F., {Dent}, W., {et~al.} 2013, \aap,
  557, A133

\bibitem[{{Dohlen} {et~al.}(2008){Dohlen}, {Langlois}, {Saisse}, {Hill},
  {Origne}, {Jacquet}, {Fabron}, {Blanc}, {Llored}, {Carle}, {Moutou}, {Vigan},
  {Boccaletti}, {Carbillet}, {Mouillet}, \& {Beuzit}}]{Dohlen:2008uh}
{Dohlen}, K., {Langlois}, M., {Saisse}, M., {et~al.} 2008, in \procspie, Vol.
  7014, Ground-based and Airborne Instrumentation for Astronomy II, ed. I.~S.
  {McLean} \& M.~M. {Casali}, 70143L

\bibitem[{{Dong} {et~al.}(2015){Dong}, {Zhu}, {Rafikov}, \&
  {Stone}}]{dong_2015}
{Dong}, R., {Zhu}, Z., {Rafikov}, R.~R., \& {Stone}, J.~M. 2015, \apj, 809, L5

\bibitem[{{Dong} {et~al.}(2018){Dong}, {Liu}, {Eisner}, {Andrews}, {Fung},
  {Zhu}, {Chiang}, {Hashimoto}, {Liu}, {Casassus}, {Esposito}, {Hasegawa},
  {Muto}, {Pavlyuchenkov}, {Wilner}, {Akiyama}, {Tamura}, \&
  {Wisniewski}}]{2018ApJ...860..124D}
{Dong}, R., {Liu}, S.-y., {Eisner}, J., {et~al.} 2018, \apj, 860, 124

\bibitem[{{Dullemond} {et~al.}(2019){Dullemond}, {K{\"u}ffmeier}, {Goicovic},
  {Fukagawa}, {Oehl}, \& {Kramer}}]{dullemond2019}
{Dullemond}, C.~P., {K{\"u}ffmeier}, M., {Goicovic}, F., {et~al.} 2019, \aap,
  628, A20

\bibitem[{{Fedele} {et~al.}(2021){Fedele}, {Toci}, {Maud}, \&
  {Lodato}}]{Fedele2021}
{Fedele}, D., {Toci}, C., {Maud}, L., \& {Lodato}, G. 2021, \aap, 651, A90

\bibitem[{{Follette} {et~al.}(2017){Follette}, {Rameau}, {Dong}, {Pueyo},
  {Close}, {Duch{\^e}ne}, {Fung}, {Leonard}, {Macintosh}, {Males}, {Marois},
  {Millar-Blanchaer}, {Morzinski}, {Mullen}, {Perrin}, {Spiro}, {Wang},
  {Ammons}, {Bailey}, {Barman}, {Bulger}, {Chilcote}, {Cotten}, {De Rosa},
  {Doyon}, {Fitzgerald}, {Goodsell}, {Graham}, {Greenbaum}, {Hibon}, {Hung},
  {Ingraham}, {Kalas}, {Konopacky}, {Larkin}, {Maire}, {Marchis}, {Metchev},
  {Nielsen}, {Oppenheimer}, {Palmer}, {Patience}, {Poyneer}, {Rajan},
  {Rantakyr{\"o}}, {Savransky}, {Schneider}, {Sivaramakrishnan}, {Song},
  {Soummer}, {Thomas}, {Vega}, {Wallace}, {Ward-Duong}, {Wiktorowicz}, \&
  {Wolff}}]{Folette2017}
{Follette}, K.~B., {Rameau}, J., {Dong}, R., {et~al.} 2017, \aj, 153, 264

\bibitem[{{Gaia Collaboration} {et~al.}(2021){Gaia Collaboration}, {Brown},
  {Vallenari}, {Prusti}, {de Bruijne}, {Babusiaux}, {Biermann}, {Creevey},
  {Evans}, {Eyer}, {Hutton}, {Jansen}, {Jordi}, {Klioner}, {Lammers},
  {Lindegren}, {Luri}, {Mignard}, {Panem}, {Pourbaix}, {Randich}, {Sartoretti},
  {Soubiran}, {Walton}, {Arenou}, {Bailer-Jones}, {Bastian}, {Cropper},
  {Drimmel}, {Katz}, {Lattanzi}, {van Leeuwen}, {Bakker}, {Cacciari},
  {Casta{\~n}eda}, {De Angeli}, {Ducourant}, {Fabricius}, {Fouesneau},
  {Fr{\'e}mat}, {Guerra}, {Guerrier}, {Guiraud}, {Jean-Antoine Piccolo},
  {Masana}, {Messineo}, {Mowlavi}, {Nicolas}, {Nienartowicz}, {Pailler},
  {Panuzzo}, {Riclet}, {Roux}, {Seabroke}, {Sordo}, {Tanga}, {Th{\'e}venin},
  {Gracia-Abril}, {Portell}, {Teyssier}, {Altmann}, {Andrae}, {Bellas-Velidis},
  {Benson}, {Berthier}, {Blomme}, {Brugaletta}, {Burgess}, {Busso}, {Carry},
  {Cellino}, {Cheek}, {Clementini}, {Damerdji}, {Davidson}, {Delchambre},
  {Dell'Oro}, {Fern{\'a}ndez-Hern{\'a}ndez}, {Galluccio}, {Garc{\'\i}a-Lario},
  {Garcia-Reinaldos}, {Gonz{\'a}lez-N{\'u}{\~n}ez}, {Gosset}, {Haigron},
  {Halbwachs}, {Hambly}, {Harrison}, {Hatzidimitriou}, {Heiter},
  {Hern{\'a}ndez}, {Hestroffer}, {Hodgkin}, {Holl}, {Jan{\ss}en}, {Jevardat de
  Fombelle}, {Jordan}, {Krone-Martins}, {Lanzafame}, {L{\"o}ffler}, {Lorca},
  {Manteiga}, {Marchal}, {Marrese}, {Moitinho}, {Mora}, {Muinonen}, {Osborne},
  {Pancino}, {Pauwels}, {Petit}, {Recio-Blanco}, {Richards}, {Riello},
  {Rimoldini}, {Robin}, {Roegiers}, {Rybizki}, {Sarro}, {Siopis}, {Smith},
  {Sozzetti}, {Ulla}, {Utrilla}, {van Leeuwen}, {van Reeven}, {Abbas}, {Abreu
  Aramburu}, {Accart}, {Aerts}, {Aguado}, {Ajaj}, {Altavilla}, {{\'A}lvarez},
  {{\'A}lvarez Cid-Fuentes}, {Alves}, {Anderson}, {Anglada Varela}, {Antoja},
  {Audard}, {Baines}, {Baker}, {Balaguer-N{\'u}{\~n}ez}, {Balbinot}, {Balog},
  {Barache}, {Barbato}, {Barros}, {Barstow}, {Bartolom{\'e}}, {Bassilana},
  {Bauchet}, {Baudesson-Stella}, {Becciani}, {Bellazzini}, {Bernet}, {Bertone},
  {Bianchi}, {Blanco-Cuaresma}, {Boch}, {Bombrun}, {Bossini}, {Bouquillon},
  {Bragaglia}, {Bramante}, {Breedt}, {Bressan}, {Brouillet}, {Bucciarelli},
  {Burlacu}, {Busonero}, {Butkevich}, {Buzzi}, {Caffau}, {Cancelliere},
  {C{\'a}novas}, {Cantat-Gaudin}, {Carballo}, {Carlucci}, {Carnerero},
  {Carrasco}, {Casamiquela}, {Castellani}, {Castro-Ginard}, {Castro Sampol},
  {Chaoul}, {Charlot}, {Chemin}, {Chiavassa}, {Cioni}, {Comoretto}, {Cooper},
  {Cornez}, {Cowell}, {Crifo}, {Crosta}, {Crowley}, {Dafonte}, {Dapergolas},
  {David}, {David}, {de Laverny}, {De Luise}, {De March}, {De Ridder}, {de
  Souza}, {de Teodoro}, {de Torres}, {del Peloso}, {del Pozo}, {Delbo},
  {Delgado}, {Delgado}, {Delisle}, {Di Matteo}, {Diakite}, {Diener},
  {Distefano}, {Dolding}, {Eappachen}, {Edvardsson}, {Enke}, {Esquej}, {Fabre},
  {Fabrizio}, {Faigler}, {Fedorets}, {Fernique}, {Fienga}, {Figueras},
  {Fouron}, {Fragkoudi}, {Fraile}, {Franke}, {Gai}, {Garabato},
  {Garcia-Gutierrez}, {Garc{\'\i}a-Torres}, {Garofalo}, {Gavras}, {Gerlach},
  {Geyer}, {Giacobbe}, {Gilmore}, {Girona}, {Giuffrida}, {Gomel}, {Gomez},
  {Gonzalez-Santamaria}, {Gonz{\'a}lez-Vidal}, {Granvik},
  {Guti{\'e}rrez-S{\'a}nchez}, {Guy}, {Hauser}, {Haywood}, {Helmi}, {Hidalgo},
  {Hilger}, {H{\l}adczuk}, {Hobbs}, {Holland}, {Huckle}, {Jasniewicz},
  {Jonker}, {Juaristi Campillo}, {Julbe}, {Karbevska}, {Kervella}, {Khanna},
  {Kochoska}, {Kontizas}, {Kordopatis}, {Korn}, {Kostrzewa-Rutkowska},
  {Kruszy{\'n}ska}, {Lambert}, {Lanza}, {Lasne}, {Le Campion}, {Le Fustec},
  {Lebreton}, {Lebzelter}, {Leccia}, {Leclerc}, {Lecoeur-Taibi}, {Liao},
  {Licata}, {Lindstr{\o}m}, {Lister}, {Livanou}, {Lobel}, {Madrero Pardo},
  {Managau}, {Mann}, {Marchant}, {Marconi}, {Marcos Santos}, {Marinoni},
  {Marocco}, {Marshall}, {Martin Polo}, {Mart{\'\i}n-Fleitas}, {Masip},
  {Massari}, {Mastrobuono-Battisti}, {Mazeh}, {McMillan}, {Messina},
  {Michalik}, {Millar}, {Mints}, {Molina}, {Molinaro}, {Moln{\'a}r},
  {Montegriffo}, {Mor}, {Morbidelli}, {Morel}, {Morris}, {Mulone}, {Munoz},
  {Muraveva}, {Murphy}, {Musella}, {Noval}, {Ord{\'e}novic}, {Orr{\`u}},
  {Osinde}, {Pagani}, {Pagano}, {Palaversa}, {Palicio}, {Panahi}, {Pawlak},
  {Pe{\~n}alosa Esteller}, {Penttil{\"a}}, {Piersimoni}, {Pineau}, {Plachy},
  {Plum}, {Poggio}, {Poretti}, {Poujoulet}, {Pr{\v{s}}a}, {Pulone}, {Racero},
  {Ragaini}, {Rainer}, {Raiteri}, {Rambaux}, {Ramos}, {Ramos-Lerate}, {Re
  Fiorentin}, {Regibo}, {Reyl{\'e}}, {Ripepi}, {Riva}, {Rixon}, {Robichon},
  {Robin}, {Roelens}, {Rohrbasser}, {Romero-G{\'o}mez}, {Rowell}, {Royer},
  {Rybicki}, {Sadowski}, {Sagrist{\`a} Sell{\'e}s}, {Sahlmann}, {Salgado},
  {Salguero}, {Samaras}, {Sanchez Gimenez}, {Sanna}, {Santove{\~n}a},
  {Sarasso}, {Schultheis}, {Sciacca}, {Segol}, {Segovia}, {S{\'e}gransan},
  {Semeux}, {Shahaf}, {Siddiqui}, {Siebert}, {Siltala}, {Slezak}, {Smart},
  {Solano}, {Solitro}, {Souami}, {Souchay}, {Spagna}, {Spoto}, {Steele},
  {Steidelm{\"u}ller}, {Stephenson}, {S{\"u}veges}, {Szabados}, {Szegedi-Elek},
  {Taris}, {Tauran}, {Taylor}, {Teixeira}, {Thuillot}, {Tonello}, {Torra},
  {Torra}, {Turon}, {Unger}, {Vaillant}, {van Dillen}, {Vanel}, {Vecchiato},
  {Viala}, {Vicente}, {Voutsinas}, {Weiler}, {Wevers}, {Wyrzykowski}, {Yoldas},
  {Yvard}, {Zhao}, {Zorec}, {Zucker}, {Zurbach}, \& {Zwitter}}]{gaia_dr3}
{Gaia Collaboration}, {Brown}, A.~G.~A., {Vallenari}, A., {et~al.} 2021, \aap,
  649, A1

\bibitem[{{Garg} {et~al.}(2021){Garg}, {Pinte}, {Christiaens}, {Price},
  {Lazendic}, {Boehler}, {Casassus}, {Marino}, {Perez}, \& {Zuleta}}]{Garg2021}
{Garg}, H., {Pinte}, C., {Christiaens}, V., {et~al.} 2021, \mnras, 504, 782

\bibitem[{{Garufi} {et~al.}(2016){Garufi}, {Quanz}, {Schmid}, {Mulders},
  {Avenhaus}, {Boccaletti}, {Ginski}, {Langlois}, {Stolker}, {Augereau},
  {Benisty}, {Lopez}, {Dominik}, {Gratton}, {Henning}, {Janson}, {M{\'e}nard},
  {Meyer}, {Pinte}, {Sissa}, {Vigan}, {Zurlo}, {Bazzon}, {Buenzli}, {Bonnefoy},
  {Brandner}, {Chauvin}, {Cheetham}, {Cudel}, {Desidera}, {Feldt}, {Galicher},
  {Kasper}, {Lagrange}, {Lannier}, {Maire}, {Mesa}, {Mouillet}, {Peretti},
  {Perrot}, {Salter}, \& {Wildi}}]{2016A&A...588A...8G}
{Garufi}, A., {Quanz}, S.~P., {Schmid}, H.~M., {et~al.} 2016, \aap, 588, A8

\bibitem[{{Gravity Collaboration} {et~al.}(2019){Gravity Collaboration},
  {Perraut}, {Labadie}, {Lazareff}, {Klarmann}, {Segura-Cox}, {Benisty},
  {Bouvier}, {Brandner}, {Caratti O Garatti}, {Caselli}, {Dougados}, {Garcia},
  {Garcia-Lopez}, {Kendrew}, {Koutoulaki}, {Kervella}, {Lin}, {Pineda},
  {Sanchez-Bermudez}, {van Dishoeck}, {Abuter}, {Amorim}, {Berger}, {Bonnet},
  {Buron}, {Cantalloube}, {Cl{\'e}net}, {Coud{\'e} Du Foresto}, {Dexter}, {de
  Zeeuw}, {Duvert}, {Eckart}, {Eisenhauer}, {Eupen}, {Gao}, {Gendron},
  {Genzel}, {Gillessen}, {Gordo}, {Grellmann}, {Haubois}, {Haussmann},
  {Henning}, {Hippler}, {Horrobin}, {Hubert}, {Jocou}, {Lacour}, {Le Bouquin},
  {L{\'e}na}, {M{\'e}rand}, {Ott}, {Paumard}, {Perrin}, {Pfuhl}, {Rabien},
  {Ray}, {Rau}, {Rousset}, {Scheithauer}, {Straub}, {Straubmeier}, {Sturm},
  {Vincent}, {Waisberg}, {Wank}, {Widmann}, {Wieprecht}, {Wiest}, {Wiezorrek},
  {Woillez}, \& {Yazici}}]{Gravity2019}
{Gravity Collaboration}, {Perraut}, K., {Labadie}, L., {et~al.} 2019, \aap,
  632, A53

\bibitem[{{Hirsh} {et~al.}(2020){Hirsh}, {Price}, {Gonzalez},
  {Ubeira-Gabellini}, \& {Ragusa}}]{hirsh2020}
{Hirsh}, K., {Price}, D.~J., {Gonzalez}, J.-F., {Ubeira-Gabellini}, M.~G., \&
  {Ragusa}, E. 2020, \mnras, 498, 2936

\bibitem[{{Jorsater} \& {van Moorsel}(1995)}]{jvm1995}
{Jorsater}, S., \& {van Moorsel}, G.~A. 1995, \aj, 110, 2037

\bibitem[{{Kraus} {et~al.}(2017){Kraus}, {Kreplin}, {Fukugawa}, {Muto},
  {Sitko}, {Young}, {Bate}, {Grady}, {Harries}, {Monnier}, {Willson}, \&
  {Wisniewski}}]{2017ApJ...848L..11K}
{Kraus}, S., {Kreplin}, A., {Fukugawa}, M., {et~al.} 2017, \apj, 848, L11

\bibitem[{{Law} {et~al.}(2021){Law}, {Teague}, {Loomis}, {Bae}, {{\"O}berg},
  {Czekala}, {Andrews}, {Aikawa}, {Alarc{\'o}n}, {Bergin}, {Bergner}, {Booth},
  {Bosman}, {Calahan}, {Cataldi}, {Cleeves}, {Furuya}, {Guzm{\'a}n}, {Huang},
  {Ilee}, {Le Gal}, {Liu}, {Long}, {M{\'e}nard}, {Nomura}, {P{\'e}rez}, {Qi},
  {Schwarz}, {Soto}, {Tsukagoshi}, {Yamato}, {van't Hoff}, {Walsh}, {Wilner},
  \& {Zhang}}]{law_2021}
{Law}, C.~J., {Teague}, R., {Loomis}, R.~A., {et~al.} 2021, \apjs, 257, 4

\bibitem[{{McMullin} {et~al.}(2007){McMullin}, {Waters}, {Schiebel}, {Young},
  \& {Golap}}]{2007ASPC..376..127M}
{McMullin}, J.~P., {Waters}, B., {Schiebel}, D., {Young}, W., \& {Golap}, K.
  2007, in Astronomical Society of the Pacific Conference Series, Vol. 376,
  Astronomical Data Analysis Software and Systems XVI, ed. R.~A. {Shaw},
  F.~{Hill}, \& D.~J. {Bell}, 127

\bibitem[{{Miley} {et~al.}(2019){Miley}, {Pani{\'c}}, {Haworth}, {Pascucci},
  {Wyatt}, {Clarke}, {Richards}, \& {Ratzka}}]{miley_2019}
{Miley}, J.~M., {Pani{\'c}}, O., {Haworth}, T.~J., {et~al.} 2019, \mnras, 485,
  739

\bibitem[{{Norfolk} {et~al.}(2021){Norfolk}, {Maddison}, {Pinte}, {van der
  Marel}, {Booth}, {Francis}, {Gonzalez}, {M{\'e}nard}, {Wright}, {van der
  Plas}, \& {Garg}}]{2021MNRAS.502.5779N}
{Norfolk}, B.~J., {Maddison}, S.~T., {Pinte}, C., {et~al.} 2021, \mnras, 502,
  5779

\bibitem[{{Paardekooper} \& {Mellema}(2004)}]{paardekooper_2004}
{Paardekooper}, S.~J., \& {Mellema}, G. 2004, \aap, 425, L9

\bibitem[{{P{\'e}rez} {et~al.}(2020){P{\'e}rez}, {Casassus}, {Hales}, {Marino},
  {Cheetham}, {Zurlo}, {Cieza}, {Dong}, {Alarc{\'o}n}, {Ben{\'\i}tez-Llambay},
  {Fomalont}, \& {Avenhaus}}]{perez_2020}
{P{\'e}rez}, S., {Casassus}, S., {Hales}, A., {et~al.} 2020, \apjl, 889, L24

\bibitem[{{Pinte} {et~al.}(2009){Pinte}, {Harries}, {Min}, {Watson},
  {Dullemond}, {Woitke}, {M{\'e}nard}, \& {Dur{\'a}n-Rojas}}]{pinte2009}
{Pinte}, C., {Harries}, T.~J., {Min}, M., {et~al.} 2009, \aap, 498, 967

\bibitem[{{Pinte} {et~al.}(2006){Pinte}, {M{\'e}nard}, {Duch{\^e}ne}, \&
  {Bastien}}]{pinte2006}
{Pinte}, C., {M{\'e}nard}, F., {Duch{\^e}ne}, G., \& {Bastien}, P. 2006, \aap,
  459, 797

\bibitem[{{Pinte} {et~al.}(2018{\natexlab{a}}){Pinte}, {Price}, {M{\'e}nard},
  {Duch{\^e}ne}, {Dent}, {Hill}, {de Gregorio-Monsalvo}, {Hales}, \&
  {Mentiplay}}]{Pinte2018}
{Pinte}, C., {Price}, D.~J., {M{\'e}nard}, F., {et~al.} 2018{\natexlab{a}},
  \apjl, 860, L13

\bibitem[{{Pinte} {et~al.}(2018{\natexlab{b}}){Pinte}, {M{\'e}nard},
  {Duch{\^e}ne}, {Hill}, {Dent}, {Woitke}, {Maret}, {van der Plas}, {Hales},
  {Kamp}, {Thi}, {de Gregorio-Monsalvo}, {Rab}, {Quanz}, {Avenhaus}, {Carmona},
  \& {Casassus}}]{pinte_2018b}
{Pinte}, C., {M{\'e}nard}, F., {Duch{\^e}ne}, G., {et~al.} 2018{\natexlab{b}},
  \aap, 609, A47

\bibitem[{{Pinte} {et~al.}(2020){Pinte}, {Price}, {M{\'e}nard}, {Duch{\^e}ne},
  {Christiaens}, {Andrews}, {Huang}, {Hill}, {van der Plas}, {Perez}, {Isella},
  {Boehler}, {Dent}, {Mentiplay}, \& {Loomis}}]{Pinte2020}
{Pinte}, C., {Price}, D.~J., {M{\'e}nard}, F., {et~al.} 2020, \apjl, 890, L9

\bibitem[{{Price} {et~al.}(2018{\natexlab{a}}){Price}, {Wurster}, {Tricco},
  {Nixon}, {Toupin}, {Pettitt}, {Chan}, {Mentiplay}, {Laibe}, {Glover},
  {Dobbs}, {Nealon}, {Liptai}, {Worpel}, {Bonnerot}, {Dipierro}, {Ballabio},
  {Ragusa}, {Federrath}, {Iaconi}, {Reichardt}, {Forgan}, {Hutchison},
  {Constantino}, {Ayliffe}, {Hirsh}, \& {Lodato}}]{Price2018b}
{Price}, D.~J., {Wurster}, J., {Tricco}, T.~S., {et~al.} 2018{\natexlab{a}},
  \pasa, 35, e031

\bibitem[{{Price} {et~al.}(2018{\natexlab{b}}){Price}, {Cuello}, {Pinte},
  {Mentiplay}, {Casassus}, {Christiaens}, {Kennedy}, {Cuadra}, {Sebastian
  Perez}, {Marino}, {Armitage}, {Zurlo}, {Juhasz}, {Ragusa}, {Laibe}, \&
  {Lodato}}]{Price2018}
{Price}, D.~J., {Cuello}, N., {Pinte}, C., {et~al.} 2018{\natexlab{b}}, \mnras,
  477, 1270

\bibitem[{{Ragusa} {et~al.}(2017){Ragusa}, {Dipierro}, {Lodato}, {Laibe}, \&
  {Price}}]{Ragusa2017}
{Ragusa}, E., {Dipierro}, G., {Lodato}, G., {Laibe}, G., \& {Price}, D.~J.
  2017, \mnras, 464, 1449

\bibitem[{{Rosotti} {et~al.}(2020){Rosotti}, {Benisty}, {Juh{\'a}sz}, {Teague},
  {Clarke}, {Dominik}, {Dullemond}, {Klaassen}, {Matr{\`a}}, \&
  {Stolker}}]{Rosotti2020}
{Rosotti}, G.~P., {Benisty}, M., {Juh{\'a}sz}, A., {et~al.} 2020, \mnras, 491,
  1335

\bibitem[{{Shakura} \& {Sunyaev}(1973)}]{shakura1973}
{Shakura}, N.~I., \& {Sunyaev}, R.~A. 1973, \aap, 24, 337

\bibitem[{{Siess} {et~al.}(2000){Siess}, {Dufour}, \& {Forestini}}]{Siess2000}
{Siess}, L., {Dufour}, E., \& {Forestini}, M. 2000, \aap, 358, 593

\bibitem[{{Sissa} {et~al.}(2018){Sissa}, {Gratton}, {Garufi}, {Rigliaco},
  {Zurlo}, {Mesa}, {Langlois}, {de Boer}, {Desidera}, {Ginski}, {Lagrange},
  {Maire}, {Vigan}, {Dima}, {Antichi}, {Baruffolo}, {Bazzon}, {Benisty},
  {Beuzit}, {Biller}, {Boccaletti}, {Bonavita}, {Bonnefoy}, {Brandner},
  {Bruno}, {Buenzli}, {Cascone}, {Chauvin}, {Cheetham}, {Claudi}, {Cudel}, {De
  Caprio}, {Dominik}, {Fantinel}, {Farisato}, {Feldt}, {Fontanive}, {Galicher},
  {Giro}, {Hagelberg}, {Incorvaia}, {Janson}, {Kasper}, {Keppler}, {Kopytova},
  {Lagadec}, {Lannier}, {Lazzoni}, {LeCoroller}, {Lessio}, {Ligi}, {Marzari},
  {Menard}, {Meyer}, {Mouillet}, {Peretti}, {Perrot}, {Potiron}, {Rouan},
  {Salasnich}, {Salter}, {Samland}, {Schmidt}, {Scuderi}, \&
  {Wildi}}]{Sissa2018}
{Sissa}, E., {Gratton}, R., {Garufi}, A., {et~al.} 2018, \aap, 619, A160

\bibitem[{{Stolker} {et~al.}(2016){Stolker}, {Dominik}, {Min}, {Garufi},
  {Mulders}, \& {Avenhaus}}]{Stolker:2016ub}
{Stolker}, T., {Dominik}, C., {Min}, M., {et~al.} 2016, \aap, 596, A70

\bibitem[{Teague(2019)}]{eddy}
Teague, R. 2019, The Journal of Open Source Software, 4, 1220.
\newblock \url{https://doi.org/10.21105/joss.01220}

\bibitem[{{Teague} {et~al.}(2018){Teague}, {Bae}, {Bergin}, {Birnstiel}, \&
  {Foreman-Mackey}}]{Teague2018}
{Teague}, R., {Bae}, J., {Bergin}, E.~A., {Birnstiel}, T., \& {Foreman-Mackey},
  D. 2018, \apjl, 860, L12

\bibitem[{Teague \& Foreman-Mackey(2018)}]{bettermoments2018}
Teague, R., \& Foreman-Mackey, D. 2018, {bettermoments: A robust method to
  measure line centroids}, vv1.0,  Zenodo, doi:10.5281/zenodo.1419754.
\newblock \url{https://doi.org/10.5281/zenodo.1419754}

\bibitem[{{Toomre}(1964)}]{toomre_1964}
{Toomre}, A. 1964, \apj, 139, 1217

\bibitem[{{van Holstein} {et~al.}(2020){van Holstein}, {Girard}, {de Boer},
  {Snik}, {Milli}, {Stam}, {Ginski}, {Mouillet}, {Wahhaj}, {Schmid}, {Keller},
  {Langlois}, {Dohlen}, {Vigan}, {Pohl}, {Carbillet}, {Fantinel}, {Maurel},
  {Orign{\'e}}, {Petit}, {Rigal}, {Sevin}, {Boccaletti}, {Le Coroller},
  {Dominik}, {Henning}, {Lagadec}, {M{\'e}nard}, {Turatto}, {Udry}, {Chauvin},
  {Feldt}, \& {Beuzit}}]{van-Holstein:2020tm}
{van Holstein}, R.~G., {Girard}, J.~H., {de Boer}, J., {et~al.} 2020, {IRDAP:
  SPHERE-IRDIS polarimetric data reduction pipeline}, ,

\bibitem[{{Vioque} {et~al.}(2018){Vioque}, {Oudmaijer}, {Baines},
  {Mendigut{\'\i}a}, \& {P{\'e}rez-Mart{\'\i}nez}}]{vioque_2018}
{Vioque}, M., {Oudmaijer}, R.~D., {Baines}, D., {Mendigut{\'\i}a}, I., \&
  {P{\'e}rez-Mart{\'\i}nez}, R. 2018, \aap, 620, A128

\bibitem[{{Walsh} {et~al.}(2017){Walsh}, {Daley}, {Facchini}, \&
  {Juh{\'a}sz}}]{2017A&A...607A.114W}
{Walsh}, C., {Daley}, C., {Facchini}, S., \& {Juh{\'a}sz}, A. 2017, \aap, 607,
  A114

\bibitem[{{Weingartner} \& {Draine}(2001)}]{weingartner2001}
{Weingartner}, J.~C., \& {Draine}, B.~T. 2001, \apj, 548, 296

\bibitem[{{Woitke} {et~al.}(2009){Woitke}, {Kamp}, \& {Thi}}]{woitke_2009}
{Woitke}, P., {Kamp}, I., \& {Thi}, W.~F. 2009, \aap, 501, 383

\end{thebibliography}
\bibliographystyle{aasjournal}

\newpage
\clearpage

\appendix

\section{ALMA Band 6 Continuum Observations} \label{sect:app_a}
\begin{figure*}[!ht]
\centering
	\includegraphics[width=0.9\textwidth]{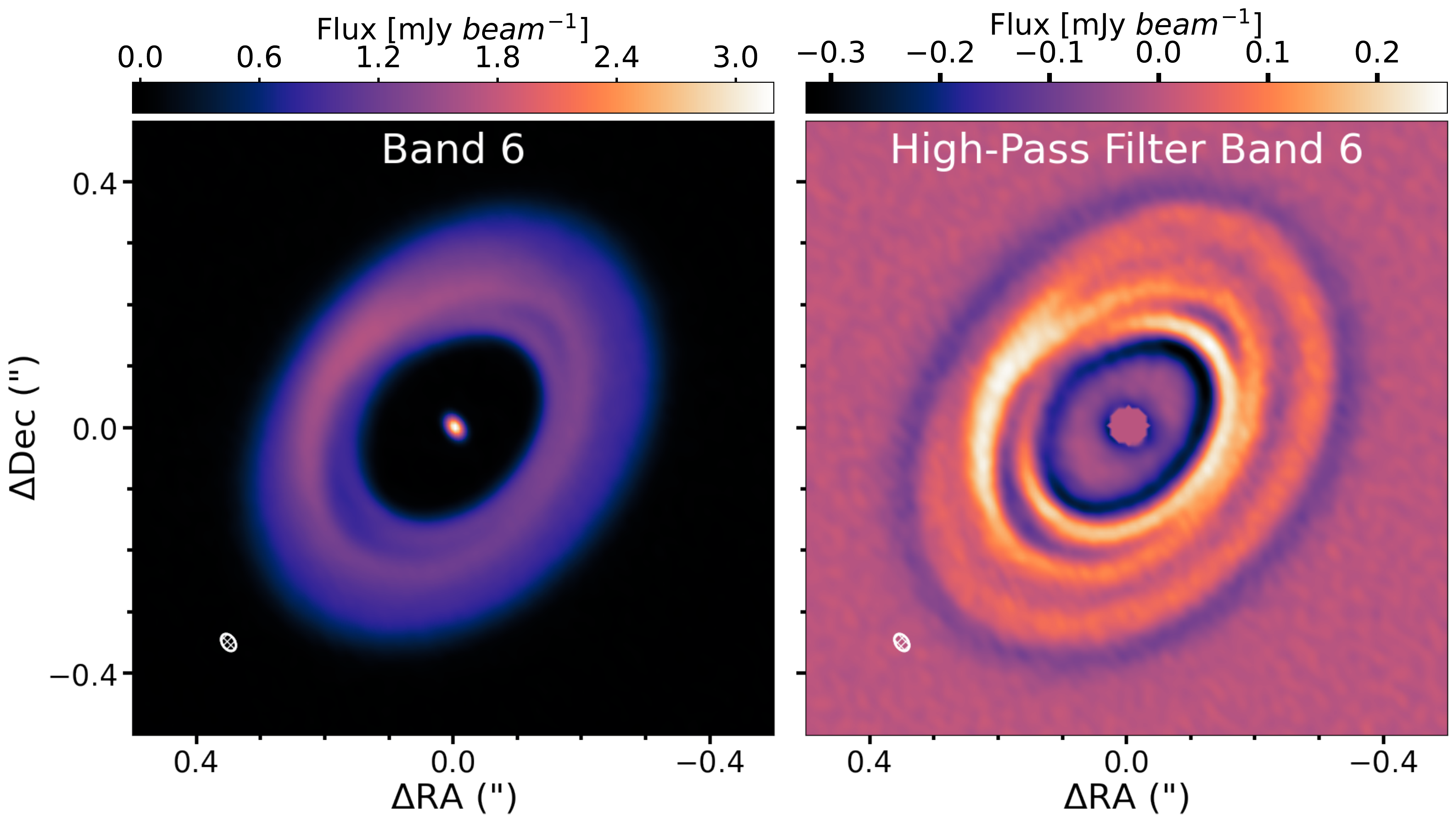}
	\caption{Left: Our Band 6 observations. Right: Our Band 6 continuum observations with a high-pass filter.}
	\label{fig:band6_dust}
\end{figure*}

Here we present our Band 6 continuum observation in Figure \ref{fig:band6_dust} (left). We apply the high-pass filter by convolving the image with an inverse Gaussian kernel, defined in Fourier space as

\begin{equation}
    K(\nu) = 1.0-\exp\left(\frac{\nu^2}{\sigma_{\nu}^2}\right),
\end{equation}

where \(\rm \nu\) is the spatial frequency and \(\rm \sigma_{\nu}\) is the width of the
filter, which we took to be 0.15 arcsec\(\rm ^{-1}\). This effectively suppresses structure on scales larger than 0.15 arcseconds. Figure \ref{fig:band6_dust} (right) shows the resulting image.

\clearpage

\section{Spirals at Large Scales} \label{sect:app_b}
\begin{figure}[!ht]
\centering
	\includegraphics[width=0.9\textwidth]{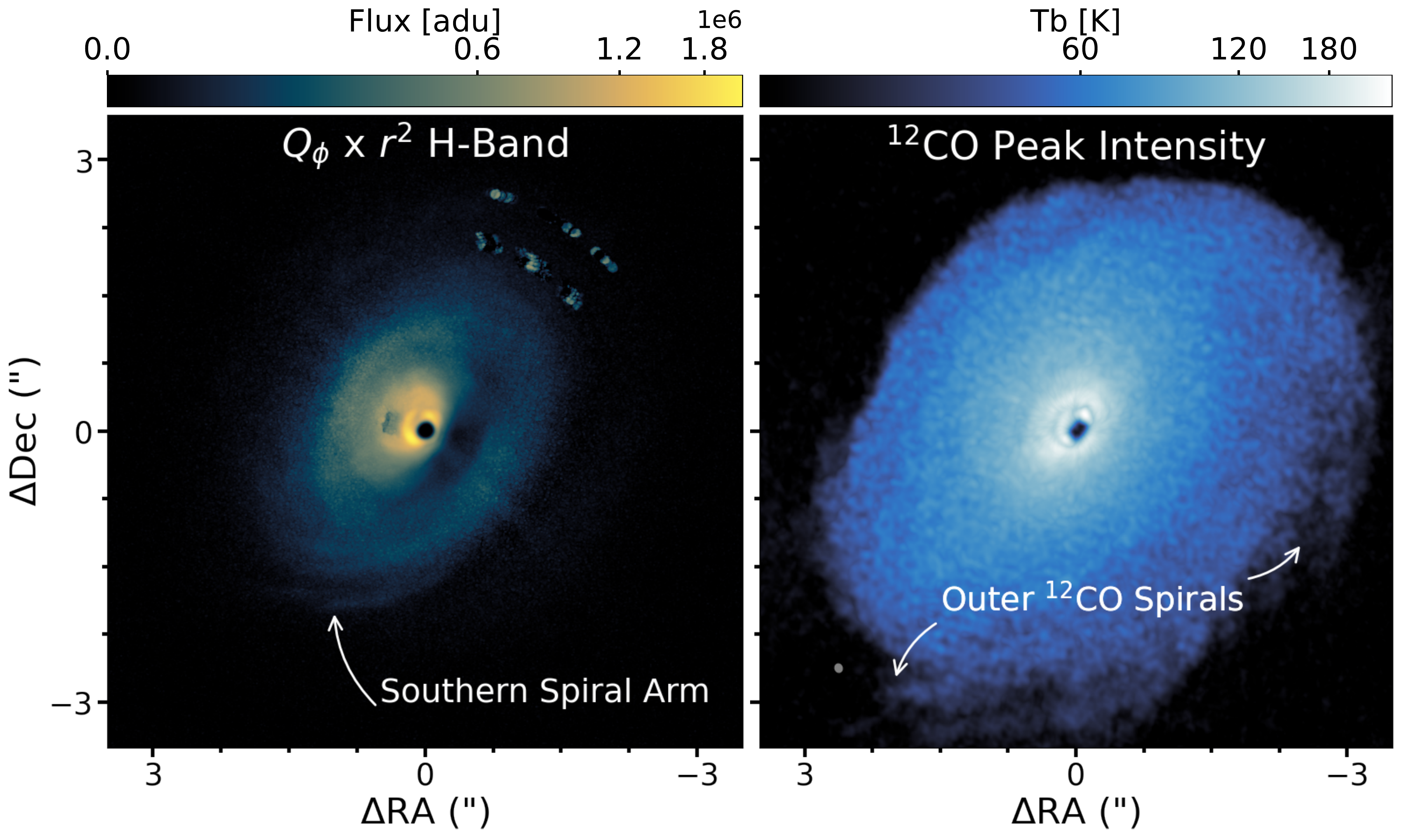}
	\caption{Left: \textit{Q\textsubscript{$\phi$}} $r^2$ $H$-band observations. Right: Peak intensity map of \(\rm ^{12}\)CO emission. A asihn scale is applied to both figures to highlight fainter outer disc emission.}
	\label{fig:qphi_m8}
\end{figure}

In Figure \ref{fig:qphi_m8} we present a large scale view of the scattered light observations and the peak intensity map of the non-continuum \(\rm ^{12}\)CO emission which highlight spiral structures at large radii. Our single binary model fails to reproduce these large scale structures suggesting that there likely exists other perturbing bodies further out in the disk. The outer companion proposed to be driving the outer dust trap \citep{Fedele2021} may also induce spirals at larger scales however, further hydrodynamical models are required to constrain the perturber(s) mass and orbit.

\clearpage

\section{Non-Keplerian Motion seen in $^{13}$CO and C$^{18}$O} \label{sect:app_e}

\begin{figure*}[!ht]
\centering
	\includegraphics[width=\textwidth]{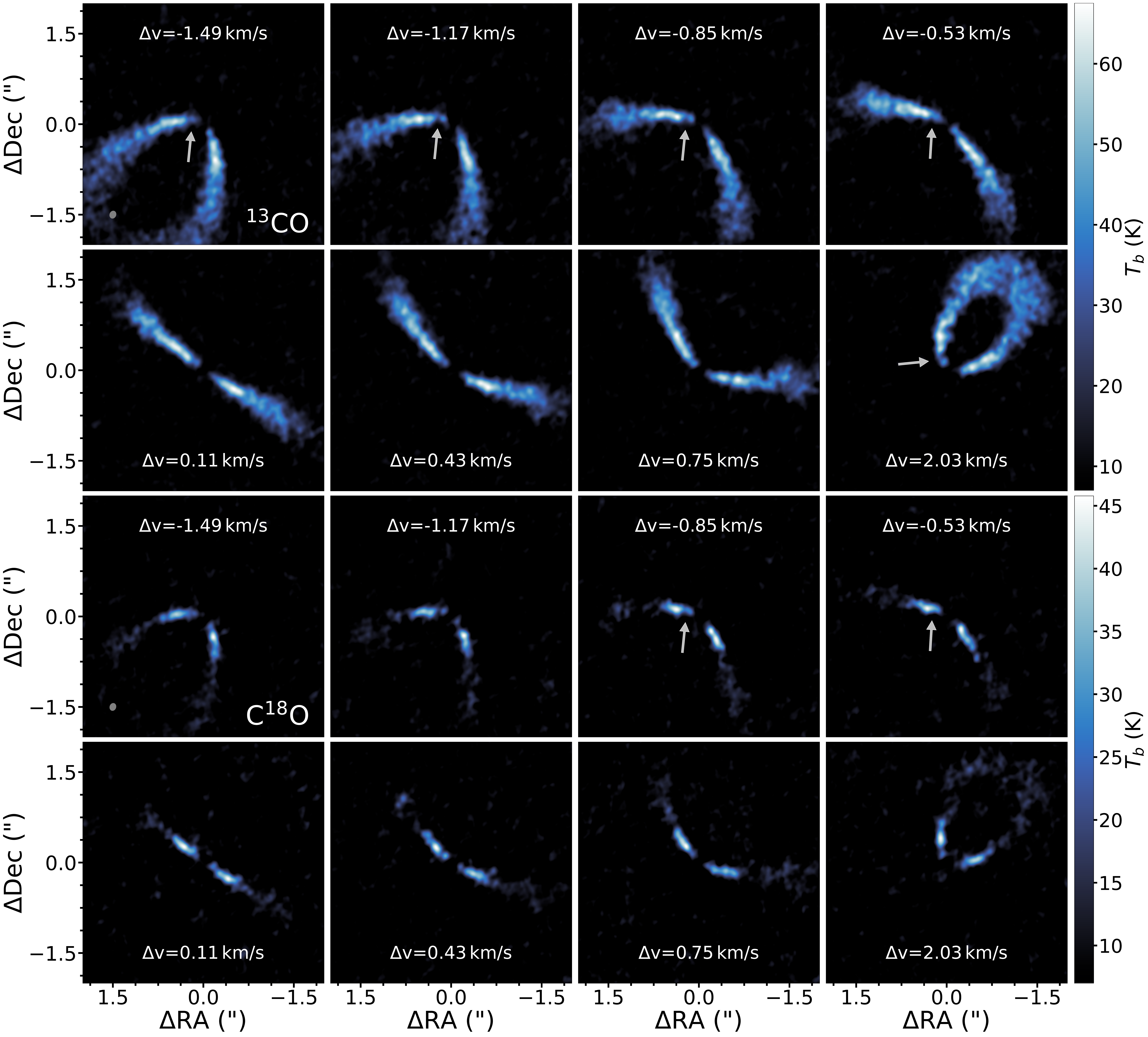}
	\caption{Large scale continuum subtracted $^{13}$CO and C$^{18}$O channel maps. We highlight a tentative inner kink in both isotopologues with a white arrow.}
	\label{fig:13_18_co_maps}
\end{figure*}

In Figure \ref{fig:13_18_co_maps} we present the continuum subtracted $^{13}$CO and C$^{18}$O channel maps. Contrary to findings by \citetalias{perez_2020} (see their Sect 3.2.2), we find that $^{13}$CO shows tentative non-Keplerian motion at small and large spatial scales, whilst the lower SNR isotopologue C$^{18}$O exhibits a tentative inner kink in only a few channels. The tentative inner kink in $^{13}$CO and C$^{18}$O spatially corresponds to the strong inner kink seen in $^{12}$CO, we highlight this feature in both isotopologues with a white arrow.

\clearpage

\section{Comparing the altitude of $^{12}$CO and scattered light emission} \label{sect:app_c}

\begin{figure*}[!ht]
\centering
	\includegraphics[width=0.65\textwidth]{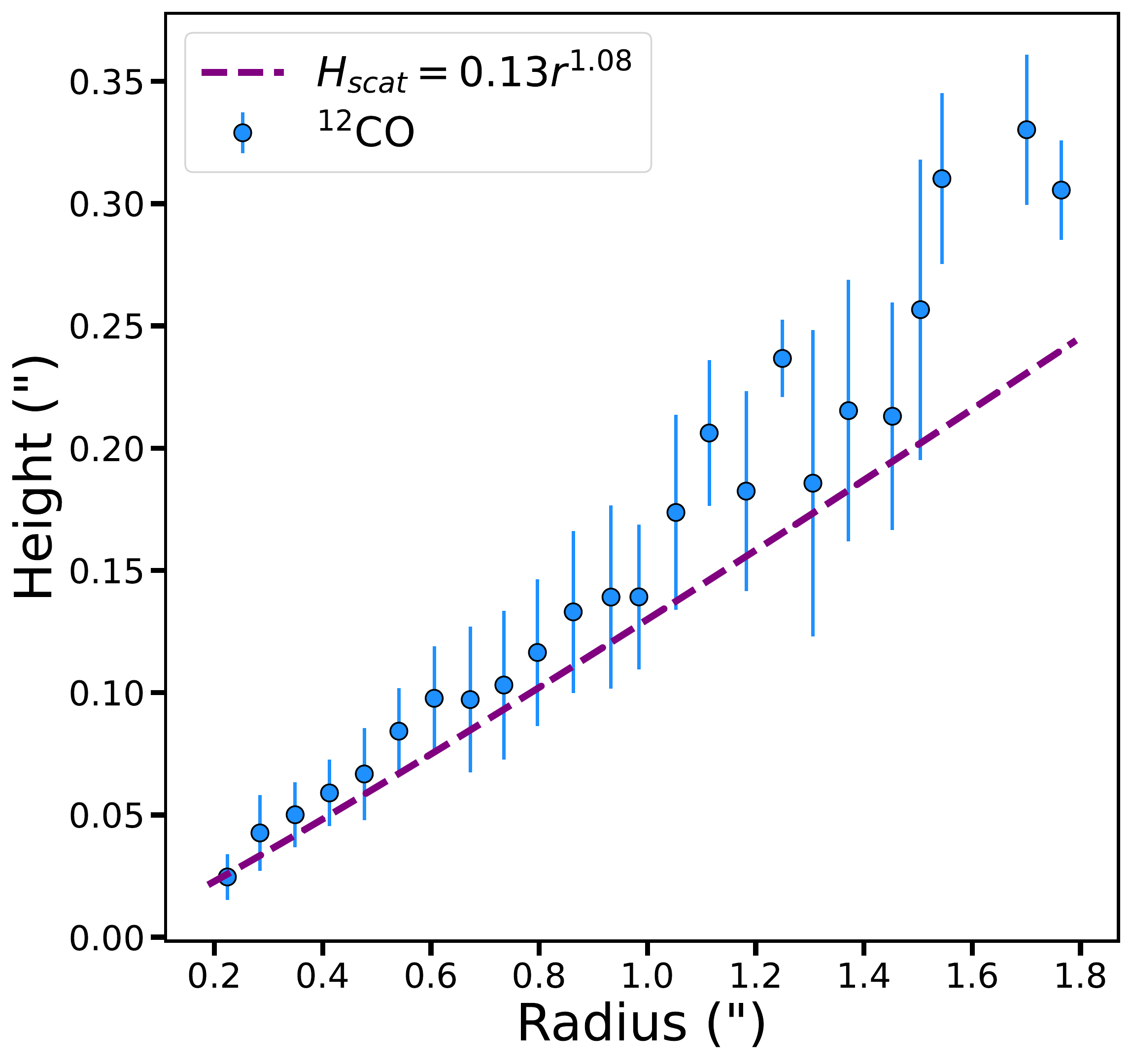}
	\caption{Disc surface height for $^{12}$CO, extracted using \textsc{co layers} \citep{pinte_2018b}, where the uncertainty is taken as the vertical deviation in each bin. The scattered light emission height is taken as \(\rm H_{scat}=0.13 \times r^{1.08}\) from \citet{Sissa2018}.}
	\label{fig:emission_height}
\end{figure*}

To compare the location of sub-structures from various disc tracers, we need to evaluated the respective emission heights to take projection effects into account. We accomplish this by tracing $^{12}$CO's upper emitting layer with the \textsc{co layers} code presented in \citet{pinte_2018b}, and over-plot the scattered light height prescription from \citet{Sissa2018}. In agreement with previous comparisons \citep{law_2021}, Figure \ref{fig:emission_height} shows that $^{12}$CO emission and the scattered light arise from a similar height in the disc.

\clearpage

\section{An inner spiral seen in SPHERE obs. and our fudicial model} \label{sect:app_d}

\begin{figure*}[!ht]
\centering
	\includegraphics[width=0.95\textwidth]{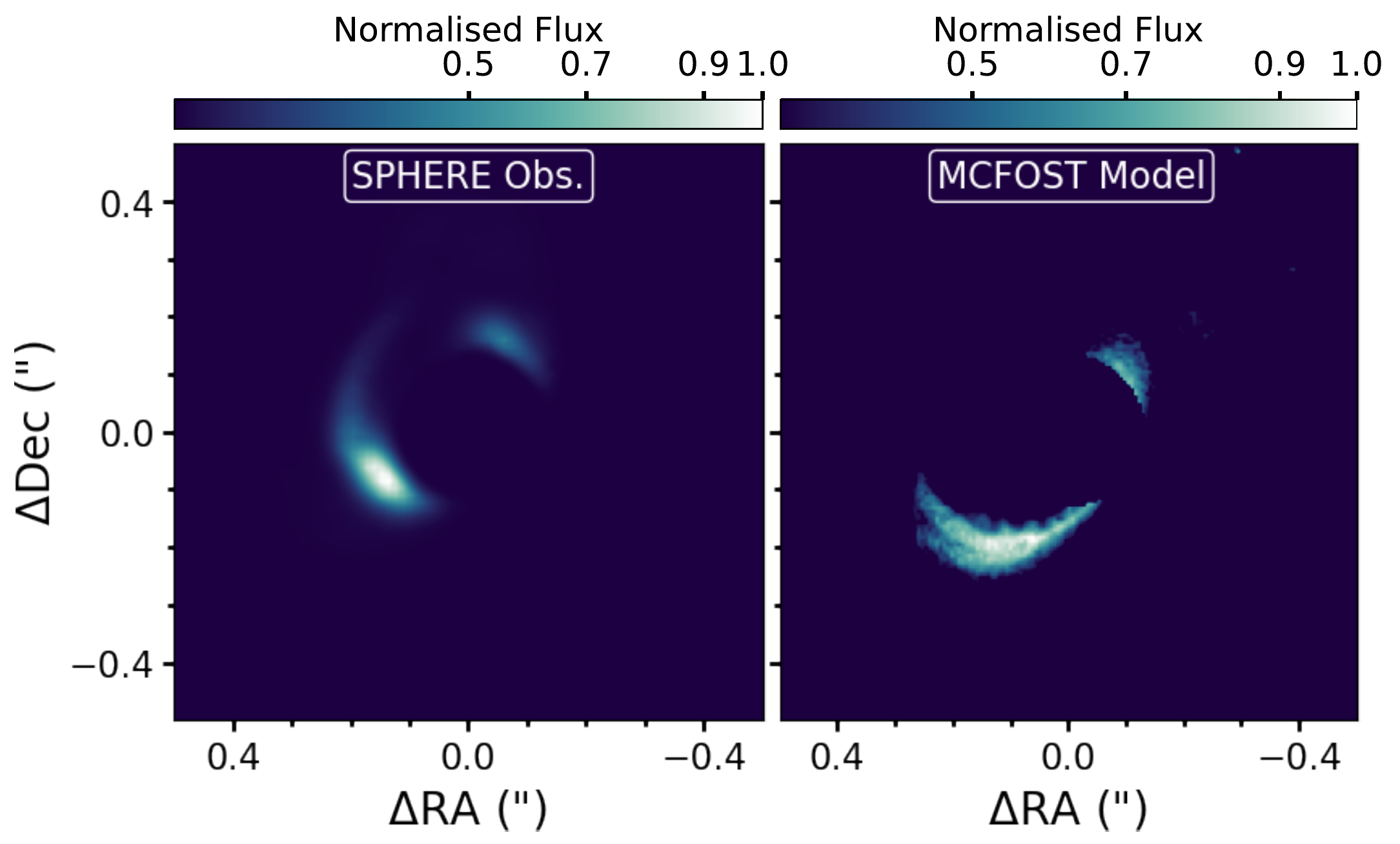}
	\caption{Left: $Q_\phi$ $r^2$-scaled \textit{H}-Band observations. Right: Synthetic $Q_\phi$ $r^2$-scaled \textit{H}-Band image produced with \textsc{mcfost}. For both panels, the flux is normalised to compare the real and synthetic data, and the colorbar is scaled to the power of 4 to solely highlight the bright inner spiral.}
	\label{fig:sphere_comp}
\end{figure*}

Figure \ref{fig:sphere_comp} presents a comparison between the SPHERE observations of the inner spiral and synthetic observations produced from our fudicial model. Similar to the observations, our synthetic $Q_\phi$ $r^2$-scaled \textit{H}-Band image shows a large inner spiral arm as well as disc emission on the opposite side of the coronagraph. However, there is an offset in the position angle between the features of the real and synthetic observations.

\end{document}